\documentclass[journal]{IEEEtran}

\usepackage{cite,citesort}
\usepackage{graphicx}
\usepackage{amsmath}
\usepackage{amssymb}
\usepackage{amsfonts}
\usepackage{algorithm}
\usepackage{algorithmic}
\usepackage{balance}
\usepackage{stfloats}
\usepackage{epstopdf}
\DeclareMathOperator*{\argmax}{argmax}

\makeatletter
\def\ScaleIfNeeded{%
\ifdim\Gin@nat@width>\linewidth \linewidth \else \Gin@nat@width
\fi } \makeatother

\newtheorem{theorem}{Theorem}

\newtheorem{corollary}{Corollary}

\begin{document}

\title{Artificial-Noise-Aided Transmission in Multi-Antenna Relay Wiretap Channels with Spatially Random Eavesdroppers}

\author{Chenxi~Liu,~\IEEEmembership{Student Member,~IEEE,}
        Nan~Yang,~\IEEEmembership{Member,~IEEE,}\\
        Robert~Malaney,~\IEEEmembership{Member,~IEEE,}
        and~Jinhong~Yuan,~\IEEEmembership{Senior Member,~IEEE}

\thanks{The work of R. Malaney and J. Yuan  was supported by the Australian Research Council Discovery Project (DP120102607). The work of N. Yang was supported by the Australian Research Council Discovery Project (DP150103905).}
\thanks{C. Liu, R. Malaney, and J. Yuan are with the School of Electrical Engineering and Telecommunications, The University of New South Wales, Sydney, NSW 2052, Australia (email:
chenxi.liu@student.unsw.edu.au; r.malaney@unsw.edu.au; j.yuan@unsw.edu.au).}
\thanks{N. Yang is with the Research School of Engineering, Australian National University, Canberra, ACT 0200, Australia (email: nan.yang@anu.edu.au).}}

\markboth{Submitted tO IEEE Transactions on Wireless Communications}{LIU
\MakeLowercase{C. Liu \textit{et al.}}: Artificial-Noise-Aided Transmission in Multi-Antenna Relay Wiretap Channels with Spatially Random Eavesdroppers}

\maketitle

\begin{abstract}
We design a new secure transmission scheme in the relay wiretap channel where a source communicates with a destination through a decode-and-forward relay in the presence of spatially random-distributed eavesdroppers. For the sake of practicality, we consider a general antenna configuration in which the source, relay, destination, and eavesdroppers are equipped with multiple antennas. In order to confuse the eavesdroppers, we assume that both the source and the relay transmit artificial noise signals in addition to information signals. We first derive a closed-form expression for the transmission outage probability and an easy-to-compute expression for the secrecy outage probability. Notably, these expressions are valid for an arbitrary number of antennas at the source, relay, and destination. We then derive simple yet valuable expressions for the asymptotic transmission outage probability and the asymptotic secrecy outage probability, which reveal the secrecy performance when the number of antennas at the source grows sufficiently large. Using our expressions, we quantify a practical performance metric, namely the secrecy throughput, under a secrecy outage probability constraint. We further determine the system and channel parameters that maximize the secrecy throughput, leading to analytical security solutions suitable for real-world deployment.
\end{abstract}


\IEEEpeerreviewmaketitle

\section{Introduction}

\PARstart{S}{ecurity} is a vital issue in wireless communication networks since data transmissions over the shared physical medium are inherently vulnerable to potential eavesdropping. Traditionally, security in wireless communication networks is realized by cryptographic techniques applied to the upper layers utilizing secret keys. The secrecy provided by such techniques is achieved under the assumption of finite computational capability at the eavesdroppers. However, this assumption cannot be easily satisfied with the rapid and continuous growth of the computational capability of modern processors, which makes the traditional cryptographic techniques increasingly weak. Moreover, the ever-expanding size of decentralized wireless networks introduces significant challenges to key distribution and management. Against this backdrop, physical layer security has been proposed as a complementary technique to traditional cryptography, due to its benefits in enhancing the secrecy level of wireless communications by direct exploiting the randomness offered by wireless channels \cite{Hong,Yang_Mag}. In seminal studies, e.g., \cite{wyner}, it was established in a single-input single-output wiretap channel that secrecy can only exist when the wiretap channel between the source and the eavesdropper is a degraded version of the main channel between the source and the legitimate receiver. This result was later generalized to the case where the main channel and the wiretap channel are independent \cite{csiszar}.

Deploying multiple antennas at the source and/or the legitimate receiver has  been shown to effectively boost the physical layer security of wiretap channels \cite{khisti,wornell,chenxi,chenxi2,nan4,Zhou10,Zhang13,nan5,nan6,nan,nan2,nan3,shihao}.
The effectiveness of multiple antennas relies on the use of secure multi-input multi-output (MIMO) techniques, such as beamforming \cite{khisti,wornell,chenxi,chenxi2,nan4}, artificial noise (AN) \cite{Zhou10,Zhang13,nan5,nan6}, and transmit antenna selection \cite{nan,nan2,nan3,shihao}. In the MIMO setting, the presence of randomly distributed eavesdroppers has been recently investigated \cite{xiangyun_2,wang_he,gio_1,gio_2,huiming}. In order to statistically characterize the secrecy performance of such scenarios, stochastic geometry and random geometric graphs are often used to model the locations of spatially random-distributed nodes. With such modeling, \cite{xiangyun_2} investigated the throughput of large-scale decentralized wireless networks with physical layer security constraints. Considering the path loss as the sole factor affecting the received signal-to-noise ratios (SNRs) at the legitimate receiver and the eavesdropper, \cite{wang_he} examined the secrecy rate in cellular networks. In \cite{gio_1} and \cite{gio_2}, the secrecy rate achieved by linear precoding was analyzed for the broadcast channel and the cellular network, respectively. In \cite{huiming}, the impact of AN was investigated.


The above works \cite{khisti,wornell,chenxi,chenxi2,nan4,Zhou10,Zhang13,nan5,nan6,nan,nan2,nan3,shihao,xiangyun_2,wang_he,gio_1,gio_2,huiming} examine physical layer security in point-to-point MIMO systems. Cooperative relaying, on the other hand, is another promising and widely-adopted technique that efficiently improves the coverage and reliability of wireless networks \cite{Laneman,Bassily}. In order to enhance physical layer security in relay wiretap channels, a variety of approaches have been investigated such as cooperative beamforming \cite{dong,huang,xiaoming_1,xiaoming_2}, relay selection \cite{yulong,hanzhu}, and cooperative jamming \cite{lai,chenxi4}. However, a common limitation of \cite{dong,huang,xiaoming_1,xiaoming_2,yulong,hanzhu,lai,chenxi4} is that they only considered fixed locations of eavesdroppers. This leaves open the problem of designing relay-aided secure transmission schemes for the scenario where the locations of eavesdroppers are spatially randomly distributed.




In this work we design a new relay-aided secure transmission for the relay wiretap channel. In such a channel, the communication between the source and the destination is aided by a decode-and-forward (DF) relay and overheard by multiple spatially random-distributed eavesdroppers. We focus on the general scenario where the source, the relay, the destination, and the eavesdroppers are equipped with multiple antennas, which stands as a major advancement over the previous studies on securing the relay wiretap channel \cite{dong,huang,xiaoming_1,xiaoming_2,yulong,hanzhu,lai,chenxi4}. In order to confuse the eavesdroppers, we assume that in the secure transmission the source and the relay transmit AN signals together with information signals in the first hop and the second hop, respectively\footnote{An initial study of a much simpler system model is given in \cite{chenxi5} where the relay, the destination, and the eavesdroppers are all equipped with a single antenna and AN signals are transmitted by the source in the first hop only. This simplified system configuration allowed for analytical tractability at the expense of significant sub-optimality.}. The contributions made by this work are summarized as follows:
\begin{enumerate}
\item
We derive a closed-form expression for the transmission outage probability and an easy-to-compute expression for the secrecy outage probability. Notably, both expressions are independent of realizations of channels and valid for an arbitrary number of antennas at the source, relay, and destination. Moreover, these expressions serve as the key results that enable us to explicitly characterize the secrecy throughput of the considered relay wiretap channels.
\item
We derive simple yet valuable expressions for the asymptotic transmission outage probability and the asymptotic secrecy outage probability. These expressions quantify the secrecy performance in the regime where the number of antennas at the source becomes sufficiently large. Based on our analysis, we find that the asymptotic transmission outage probability is determined by the average SNR of the relay-destination channel only. We also find that the asymptotic secrecy outage probability approaches a certain value which is independent of the number of antennas at the source.
\item
We determine the transmission parameters, i.e., the wiretap code rates and the power allocation factors, that maximize the secrecy throughput of the considered relay wiretap channels under a secrecy outage probability constraint. Moreover, we demonstrate the effectiveness of the determined transmission parameters on maximizing the secrecy throughput. Furthermore, we evaluate the impact of the system parameters, e.g., the number of antennas and the density of eavesdroppers, on the secrecy throughput.
\end{enumerate}

Beyond the above contributions, we provide some pivotal insights into the practical design of secure transmission. First, we show that the AN signals from the source play a more dominant role in securing the transmission in the considered relay wiretap channels than the AN signals from the relay. Second, we show that adding extra antennas at the source significantly increases the maximum secrecy throughput, but does not decrease the secrecy outage probability always. Third, we find that in order to achieve the maximum secrecy throughput, the source needs to allocate a higher power to AN signals whereas the relay needs to allocate a lower power to AN signals when the antenna number at the source increases. Fourth, we find that the maximum secrecy throughput increases when the eavesdroppers are more dispersed.

The rest of the paper is organized as follows. Section \ref{sec:system_model} describes the relay wiretap channel considered in the paper. In Section \ref{sec:analysis}, we derive expressions for the outage probabilities of the considered relay wiretap channel. The characterization and maximization of the secrecy throughput are also provided in Section \ref{sec:analysis}. Numerical results and related discussions are presented in Section \ref{sec:numerical}. Finally, Section \ref{sec:conclusion} draws conclusions.

{\em Notations}: Column vectors (matrices) are denoted by boldface lower (upper) case letters. Conjugate transpose is denoted by $\left(\cdot\right)^H$. The determinant of a matrix is denoted by $\det\left(\cdot\right)$. Complex Gaussian distribution is denoted by $\mathcal{CN}$.  A zero matrix and an identity matrix of appropriate dimension are denoted by $\mathbf{0}$ and $\mathbf{I}$, respectively. Statistical expectation is denoted by $\mathbb{E}$. The Frobenius norm of a vector or a matrix is denoted by $\|\cdot\|$.

\section{Multi-Antenna Relay Wiretap Channel}\label{sec:system_model}

\begin{figure}[!t]
\begin{center}{\includegraphics[width=0.9\columnwidth]{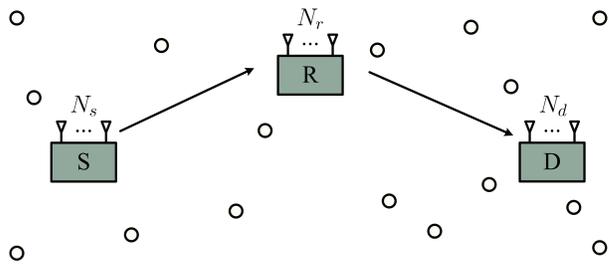}}
\caption{Illustration of a relay wiretap channel in the presence of spatially
random multi-antenna eavesdroppers.}\label{system_model}
\end{center}
\end{figure}

We consider a relay wiretap channel, as depicted in Fig. \ref{system_model}, where a source (S) communicates with a destination (D) with the aid of a relay (R) in the presence of multiple spatially random eavesdroppers. In this channel, the source, the relay, the destination, and
each eavesdropper are equipped with $N_s$, $N_r$, $N_d$, and $N_e$ antennas, respectively. We denote $\mathbf{H}_{sr}$ as the $N_r\times N_s$ channel matrix from the source to the relay and denote $\mathbf{H}_{rd}$ as the $N_d\times N_r$ channel matrix from the relay to the destination. We consider that all the channels are subject to independent and identically distributed (i.i.d) Rayleigh fading. We also consider a quasi-static block fading environment in which all
the channel coefficients remain the same within one time slot. We assume that the channel state information (CSI) between the source and the relay and the CSI between the relay and the destination are known at the source, while the CSI from the eavesdroppers is not known. We also assume that $N_s>N_e$, mimicking the case where the source is a base station (BS) with a large number of antennas, while the eavesdroppers are mobile users with a limited number of antennas. We further assume that the destination is located remotely away from the source such that the destination cannot receive signals from the source directly. All the nodes operate in a half-duplex mode such that each node cannot transmit and receive simultaneously. We denote $d_{sr}$ and $d_{rd}$ as the source-relay distance and the relay-destination distance, respectively, and denote $\eta$ as the path loss exponent. The locations of the eavesdroppers are modeled as a homogeneous Poisson Point Process (PPP) $\Phi$ with density $\lambda$ \cite{xiangyun_2,wang_he,gio_1,gio_2,huiming}, which represents the case where the eavesdroppers are mobile users in a decentralized network \cite{weber}. We clarify that the source, the relay, and the destination do not belong to $\Phi$.

\subsection{Transmission of Artificial Noise Signals}

We now detail the transmission scheme between the source and the destination. In this scheme we assume that both the source and the relay transmit AN signals together with the information signals. This scheme utilizes two time slots. In the first time slot, the source transmits information signals and AN signals to the relay, referred to as the first hop transmission. We assume that the relay adopts maximum-ratio combining (MRC) \cite{dighe,MRC,mathew} to process the received signals in order to maximize the received SNR. In the second time slot, the DF relay transmits the re-encoded signals and AN signals to the destination, referred to as the second hop transmission. We assume that the destination also adopts MRC to process the received signals. In the second time slot, it is assumed that the source transmits AN signals to further confuse the eavesdropper. We clarify that both the first hop transmission and the second hop transmission are overheard by the eavesdroppers.

In the first hop transmission, the signal transmitted by the source is given by
\begin{align}\label{x_s}
\mathbf{x}_{\textrm{S}} = \mathbf{W}_1\mathbf{t}_1,
\end{align}
where $\mathbf{W}_1$ denotes the $N_s\times N_s$ beamforming matrix at the source and $\mathbf{t}_1$ denotes the combination of the information signal and the AN signal at the source. To transmit $\mathbf{x}_{\textrm{S}}$, we first design $\mathbf{W}_1$ as
\begin{align}\label{AN_beamforming}
\mathbf{W}_1=\begin{bmatrix}
\mathbf{w}_{\textrm{S}}&\mathbf{W}_{\textrm{SAN}}
\end{bmatrix},
\end{align}
where $\mathbf{w}_{\textrm{S}}$ is used to transmit the information signal at the source and $\mathbf{W}_{\textrm{SAN}}$ is used to transmit the AN signal at the source. The aim of $\mathbf{W}_1$ is to degrade the quality of the received signals at the eavesdroppers. By transmitting AN signals through $\mathbf{W}_1$, together with the fact that the relay adopts MRC to process the received signals from the source, we ensure that the quality of the received signals at the relay is free from AN interference. In designing $\mathbf{W}_1$, we choose $\mathbf{w}_{\textrm{S}}$ as the eigenvector corresponding to the largest non-zero eigenvalue of $\mathbf{H}_{sr}^H\mathbf{H}_{sr}$, denoted by $\lambda_{\max}^{sr}$. We then choose $\mathbf{W}_{\textrm{SAN}}$ as the remaining $N_s-1$ eigenvectors of $\mathbf{H}_{sr}^H\mathbf{H}_{sr}$. Such design ensures that $\mathbf{W}_1$ is a unitary matrix. We then design $\mathbf{t}$ as
\begin{align}\label{t_s}
\mathbf{t}_1=\begin{bmatrix} t_{\textrm{S}}\\ \mathbf{t}_{\textrm{SAN}}
\end{bmatrix},
\end{align}
where $t_{\textrm{S}}$ denotes the information signal at the source and $\mathbf{t}_{\textrm{SAN}}$ is an $\left(N_s-1\right)\times 1$ vector of the AN signal at the source. We define $\beta_s$, $0<\beta_s\leq 1$, as the fraction of the power allocated to the information signal at the source. As such, we have $\mathbb{E}\left[|t_{\textrm{S}}|^2\right]=\beta_s$ and $\mathbb{E}\left[\mathbf{t}_{\textrm{SAN}}\mathbf{t}_{\textrm{SAN}}^H\right]=\frac{1-\beta_s}{N_s-1}\mathbf{I}_{N_s-1}$.
Based on \eqref{x_s}, \eqref{AN_beamforming}, and \eqref{t_s}, the received signal at the relay in the first hop transmission is expressed as
\begin{align}\label{signal_relay}
y_r=\sqrt{P_sd_{sr}^{-\eta}}\mathbf{H}_{sr}\left(\mathbf{w}_{\textrm{S}}t_{\textrm{S}}
+\mathbf{W}_{\textrm{SAN}}\mathbf{t}_{\textrm{SAN}}\right)+\mathbf{n}_r,
\end{align}
where $P_s$ denotes the transmit power at the source and $\mathbf{n}_r$ denotes the thermal noise at the relay, the elements of which are assumed to be i.i.d complex Gaussian random variables with zero mean and variance $\sigma_r^2$, i.e., $\mathbf{n}_r\sim\mathcal{CN}\left(\mathbf{0}_{N_r},\sigma_r^2\mathbf{I}_{N_r}\right)$. We note that AN signals in \eqref{signal_relay} can be canceled at the relay by applying MRC.

We next express the received signal at a typical eavesdropper located at $i$, $i\in\Phi$, in the first hop transmission as
\begin{align}\label{signal_i}
\mathbf{y}_i^{(1)}=\sqrt{P_sd_{si}^{-\eta}}\mathbf{H}_{si}\left(\mathbf{w}_{\textrm{S}}t_{\textrm{S}}
+\mathbf{W}_{\textrm{SAN}}\mathbf{t}_{\textrm{SAN}}\right)+\mathbf{n}_{i1},
\end{align}
where $\mathbf{H}_{si}$ denotes the $N_e\times N_s$ channel matrix from
the source to the typical eavesdropper located at $i$, $d_{si}$
denotes the distance between the source and the typical eavesdropper
located at $i$, and $\mathbf{n}_{i1}$ denotes the thermal noise vector at the
typical eavesdropper located at $i$, the elements of which are assumed to be i.i.d complex
Gaussian random variables with zero mean and variance $\sigma_{i1}^2$, i.e.,
$\mathbf{n}_{i1}\sim\mathcal{CN}\left(\mathbf{0}_{N_e},\sigma_{i1}^2\mathbf{I}_{N_e}\right)$.

In the second time slot, the DF relay first decodes the received signals from the source. If the received signals are successfully decoded, the relay retransmits the re-encoded signals and AN signals to the destination. The signals transmitted by the relay is given by
\begin{align}\label{x_r}
\mathbf{x}_{\textrm{R}} = \mathbf{W}_2\mathbf{t}_2,
\end{align}
where $\mathbf{W}_2$ denotes the $N_r\times N_r$ beamforming matrix at the relay and $\mathbf{t}_2$ denotes the combination of the information signal and the AN signal at the relay. Similar to $\mathbf{W}_1$ and $\mathbf{t}_1$, we design $\mathbf{W}_2$ and $\mathbf{t}_2$ as
\begin{align}\label{AN_beamforming_2}
\mathbf{W}_2=\begin{bmatrix}
\mathbf{w}_{\textrm{R}}&\mathbf{W}_{\textrm{RAN}}\end{bmatrix},
\end{align}
and
\begin{align}\label{t_r}
\mathbf{t}_2=\begin{bmatrix}
{t}_{\textrm{R}}\\\mathbf{t}_{\textrm{RAN}}\end{bmatrix},
\end{align}
respectively. In \eqref{AN_beamforming_2}, $\mathbf{w}_{\textrm{R}}$ is used to transmit the information signal at the relay and $\mathbf{W}_{\textrm{RAN}}$ is used to transmit the AN signal at the relay. In designing $\mathbf{W}_2$, we choose $\mathbf{w}_{\textrm{R}}$ as the eigenvector corresponding to the largest eigenvalue of $\mathbf{H}_{rd}^H\mathbf{H}_{rd}$, denoted by $\lambda_{\max}^{rd}$. We then choose $\mathbf{W}_{\textrm{RAN}}$ as the remaining $N_r-1$ eigenvectors of $\mathbf{H}_{rd}^H\mathbf{H}_{rd}$. This design ensures that the quality of the received signals at the destination is free from AN interference when the destination applies MRC to process the received signals. In \eqref{t_r}, $t_{\textrm{R}}$ denotes the information signal at the relay and $\mathbf{t}_{\textrm{RAN}}$ is an $(N_r-1)\times 1$ vector of the AN signals at the relay. We define $\beta_r$, $0<\beta_r\leq 1$, as the fraction of the power allocated to the information signals at the relay. As such, we have $\mathbb{E}\left[|t_{\textrm{R}}|^2\right]=\beta_r$ and $\mathbb{E}\left[\mathbf{t}_{\textrm{RAN}}\mathbf{t}_{\textrm{RAN}}^H\right]=\frac{1-\beta_r}{N_r-1}\mathbf{I}_{N_r-1}$.
According to \eqref{x_r}, \eqref{AN_beamforming_2}, and \eqref{t_r}, we express the received signal at the destination in the second hop transmission as
\begin{align}\label{signal_d}
\mathbf{y}_d=\sqrt{P_rd_{rd}^{-\eta}}\mathbf{H}_{rd}\left(\mathbf{w}_{\textrm{R}}{t}_{\textrm{R}}+\mathbf{W}_{\textrm{RAN}}\mathbf{t}_{\textrm{RAN}}\right)+\mathbf{n}_d,
\end{align}
where $P_r$ denotes the transmit power at the relay and $\mathbf{n}_d$ denotes the thermal noise at the destination, the elements of which are assumed to be i.i.d complex random variables with zero mean and variance $\sigma_d^2$, i.e., $\mathbf{n}_d\sim\mathcal{CN}\left(\mathbf{0}_{N_d},\sigma_d^2\mathbf{I}_{N_d}\right)$. We note that AN signals in \eqref{signal_d} can also be canceled at the destination by applying MRC.

In order to further confuse the eavesdroppers in the second hop transmission, we assume that the source transmits AN signals using transmit power $P_s$. We denote the AN signals from the source in the second hop transmission as $\mathbf{x}_{\textrm{AN}}$, the elements of which follow the i.i.d zero mean complex Gaussian distribution. We assume that $\mathbf{x}_{\textrm{AN}}$ has unit power such that $\mathbb{E}\left[\mathbf{x}_{\textrm{AN}}\mathbf{x}_{\textrm{AN}}^H\right]=\mathbf{I}_{N_s}/N_s$.
We next express the received signal in the second hop transmission at a typical eavesdropper located at $i$, $i\in\Phi$, as
\begin{align}\label{signal_i2}
\mathbf{y}_i^{(2)}=&\sqrt{P_rd_{ri}^{-\eta}}\mathbf{H}_{ri}\left(\mathbf{w}_{\textrm{R}}{t}_{\textrm{R}}+\mathbf{W}_{\textrm{RAN}}\mathbf{t}_{\textrm{RAN}}\right)\notag\\
&\hspace{3cm}+\sqrt{P_sd_{si}^{-\eta}}\mathbf{H}_{si}\mathbf{x}_{\textrm{AN}}+\mathbf{n}_{i2},
\end{align}
where $\mathbf{H}_{ri}$ denotes the $N_e\times N_r$ channel matrix from the relay to the typical eavesdropper located at $i$, $d_{ri}$ denotes the distance between the relay and the typical eavesdropper located at $i$, and $\mathbf{n}_{i2}$ denotes the thermal noise vector at a typical eavesdropper located at $i$, the elements of which are assumed to be i.i.d complex Gaussian random variables with zero mean and variance $\sigma_{i2}^2$, i.e., $\mathbf{n}_{i2}\sim\mathcal{CN}\left(\mathbf{0}_{N_e},\sigma_{i2}^2\mathbf{I}_{N_e}\right)$.

\subsection{Formulation of Received Signal-to-Noise Ratios}

We first focus on the equivalent instantaneous SNR at the destination. Recall that both the relay and the destination apply MRC to process received signals. We express the MRC combiner at the relay in the first hop transmission as $\mathbf{v}_r=\frac{\mathbf{w}_{\textrm{S}}^H\mathbf{H}_{sr}^{H}}{\|\mathbf{H}_{sr}\mathbf{w}_{\textrm{S}}\|}$, and express the MRC combiner at the destination in the second hop transmission as $\mathbf{v}_d=\frac{\mathbf{w}_{\textrm{R}}^H\mathbf{H}_{rd}^H}{\|\mathbf{H}_{rd}\mathbf{w}_{\textrm{R}}\|}$.
Using $\mathbf{v}_r$ and $\mathbf{v}_d$, we express the instantaneous SNR at the relay in the first hop transmission and the instantaneous SNR at the destination in the second hop transmission as
$\gamma_{sr}=\frac{\beta_s P_s}{d_{sr}^{\eta}\sigma_r^{2}}\lambda_{\max}^{sr}$ and $\gamma_{rd}=\frac{\beta_r P_r}{d_{rd}^{\eta}\sigma_d^{2}}\lambda_{\max}^{rd}$, respectively. As per the rules of the DF protocol, we express the equivalent end-to-end SNR from the source to the destination as \cite{Laneman}
\begin{align}\label{snr_d}
\Gamma_D = \min\left\{\gamma_{sr},\gamma_{rd}\right\}.
\end{align}

We now focus on the equivalent SNR at the eavesdroppers. In order to maximize the probability of successful eavesdropping, we assume that the eavesdropper utilizes the minimum mean square error (MMSE) combining to process the received signals within two time slots. As per the rules of the MMSE combining, we express the instantaneous SNR at a typical eavesdropper located at $i$ in the first hop transmission and the second hop transmission as
\begin{align}\label{snr_si}
\gamma_{si}=\beta_s{}P_sd_{si}^{-\eta}\mathbf{w}_{\textrm{S}}^H\mathbf{H}_{si}^H
\mathbf{K}_{si}^{-1}\mathbf{H}_{si}\mathbf{w}_{\textrm{S}},
\end{align} and
\begin{align}\label{snr_ri}
\gamma_{ri}=\beta_r{}P_rd_{ri}^{-\eta}\mathbf{w}_{\textrm{R}}^H\mathbf{H}_{ri}^H\mathbf{K}_{ri}^{-1}\mathbf{H}_{ri}\mathbf{w}_{\textrm{R}},
\end{align}
respectively, where
\begin{align}\label{k_si}
\mathbf{K}_{si}=\frac{1-\beta_s}{N_s-1}P_sd_{si}^{-\eta}\mathbf{H}_{si}\mathbf{W}_{\textrm{SAN}}\mathbf{W}_{\textrm{SAN}}^H\mathbf{H}_{si}^H+\sigma_{i1}^2\mathbf{I}_{N_e},
\end{align}
and
\begin{align}
\mathbf{K}_{ri}=&\frac{1-\beta_r}{N_r-1}P_rd_{ri}^{-\eta}\mathbf{H}_{ri}\mathbf{W}_{\textrm{RAN}}\mathbf{W}_{\textrm{RAN}}^H\mathbf{H}_{ri}^H\notag\\
&\hspace{2.5cm}+\frac{P_s}{N_s}d_{si}^{-\eta}\mathbf{H}_{si}\mathbf{H}_{si}^H+\sigma_{i2}\mathbf{I}_{N_e}.
\end{align}
We assume that the eavesdroppers are non-colluding, indicating that each eavesdropper decodes her own received signals from the source and the relay without cooperating with other eavesdroppers. We also assume that the source and the relay use different codebooks. As such, the transmitted signals from the source and the transmitted signals from the relay cannot be jointly processed at each eavesdropper. Based on \eqref{snr_si} and \eqref{snr_ri}, we express the equivalent SNR at the eavesdroppers as
\begin{align}\label{snr_e}
\Gamma_E=\max_{i\in\Phi}\left\{\max\left\{\gamma_{si},\gamma_{ri}\right\}\right\}.
\end{align}

\section{Secrecy Performance Analysis}\label{sec:analysis}

In this section, we analyze the secrecy performance achieved by the transmission scheme detailed in Section \ref{sec:system_model}. We first derive a closed-form expression for the transmission outage probability and an easy-to-compute expression for the secrecy outage probability, both of which are valid for an arbitrary number of antennas at the source, relay, and destination. We then derive simple yet valuable expressions for the asymptotic transmission outage probability and the asymptotic secrecy outage probability, both of which are valid for a sufficiently large number of antennas at the source, i.e., $N_s\to\infty$. We further describe in detail how the secrecy throughput of the relay wiretap channel is quantified and how the maximum secrecy throughput is obtained under a secrecy outage probability constraint.

\subsection{Preliminaries}

In this subsection, we present the statistics of $\gamma_{sr}$, $\gamma_{rd}$, $\gamma_{si}$, and $\gamma_{ri}$, which will be used to derive the outage probabilities. We first focus on the  cumulative distribution functions (CDFs) of $\gamma_{sr}$ and $\gamma_{rd}$. To this end, we introduce several new notations as follows: $u_1 = \min(N_s,N_r)$, $v_1=\max(N_s,N_r)$, $t_1=v_1-u_1$, $u_2=\min\left(N_r,N_d\right)$, $v_2=\max\left(N_r,N_d\right)$, and $t_2=v_2-u_2$.
We then obtain the CDF of $\gamma_{sr}$ as \cite{Dighe}
\begin{align}\label{cdf_1}
F_{\gamma_{sr}}\left(\gamma\right)=\frac{\det\left(\mathbf{\Xi}\left(\frac{\gamma}{\beta_s\overline{\gamma}_{sr}}\right)\right)}{\Gamma_{u_1}\left(u_1\right)\Gamma_{v_1}\left(u_1\right)},
\end{align}
where $\mathbf{\Xi}\left(\frac{\gamma}{\beta_s\overline{\gamma}_{sr}}\right)$ is a $u_1\times u_1$ matrix with $(i,j)$th entry, $\mathbf{\xi}_{ij}\left(\frac{\gamma}{\beta_s\overline{\gamma}_{sr}}\right)$, given by
\begin{align}\label{XI}
\mathbf{\xi}_{ij}\left(\frac{\gamma}{\beta_s\overline{\gamma}_{sr}}\right)=\gamma\left(g_1\left(i,j\right),\frac{\gamma}{\beta_s\overline{\gamma}_{sr}}\right).
\end{align}
In \eqref{XI}, $\gamma\left(\cdot\right)$ denotes the incomplete gamma function, defined as \cite[Eq. (8.352)]{table}
\begin{align}\label{incomplete_gamma} \gamma\left(k,x\right)=\Gamma\left(k\right)\left(1-\exp\left(-x\right)\sum_{z=0}^{k-1}\frac{x^z}{z!}\right)
\end{align}
for integer $k$,  where $\Gamma\left(\cdot\right)$ denotes the gamma function, defined as
$\Gamma\left(k\right)=(k-1)!$ for integer $k$ \cite[Eq. (8.339)]{table},
$g_1\left(i,j\right)=t_1+i+j-1$, $\overline{\gamma}_{sr}=P_sd_{sr}^{-\eta}\sigma_r^{-2}$, and
\begin{align}
\Gamma_m\left(n\right)=\prod_{i=1}^n\Gamma\left(m-i+1\right).
\end{align}
Similarly, we obtain the CDF of $\gamma_{rd}$ as
\begin{align}\label{cdf_2}
F_{\gamma_{rd}}\left(\gamma\right)=\frac{\det\left(\mathbf{\Theta}\left(\frac{\gamma}{\beta_r\overline{\gamma}_{rd}}\right)\right)}{\Gamma_{u_2}\left(u_2\right)\Gamma_{v_2}\left(u_2\right)},
\end{align}
where $\mathbf{\Theta}\left(\frac{\gamma}{\beta_r\overline{\gamma}_{rd}}\right)$ is a $u_2\times u_2$ matrix with $\left(i,j\right)$th entry, $\mathbf{\theta}_{ij}\left(\frac{\gamma}{\beta_r\overline{\gamma}_{rd}}\right)$, given by
\begin{align}
\mathbf{\theta}_{ij}\left(\frac{\gamma}{\beta_r\overline{\gamma}_{rd}}\right)=\gamma\left(g_2\left(i,j\right),\frac{\gamma}{\beta_r\overline{\gamma}_{rd}}\right),
\end{align}
where $g_2\left(i,j\right)=t_2+i+j-1$, and $\overline{\gamma}_{rd}=P_rd_{rd}^{-\eta}\sigma_d^{-2}$.

With the aid of \cite{smith}, we express the CDF of $\gamma_{si}$ as
\begin{align}\label{cdf_3}
F_{\gamma_{si}}\left(\gamma\right)=&1-\frac{\exp\left(-\frac{\gamma}{\beta_s\overline{\gamma}_{si}}\right)}{\left(1+\kappa_1\gamma\right)^{N_s-1}}\sum_{p=1}^{N_e}\frac{1}{\Gamma\left(p\right)}\left(\frac{\gamma}{\beta_s\overline{\gamma}_{si}}\right)^{p-1}\notag\\
&\times\sum_{q=0}^{N_e-p}{N_s-1\choose q}\left(\kappa_1\gamma\right)^q
\end{align}
where $\overline{\gamma}_{si}=P_sd_{si}^{-\eta}\sigma_{i1}^{-2}$ and $\kappa_1 = \dfrac{1-\beta_s}{\beta_s\left(N_s-1\right)}$, and express the CDF of $\gamma_{ri}$ as
\begin{align}\label{cdf_4}
&F_{\gamma_{ri}}\left(\gamma\right)\notag\\=&1-\frac{\exp\left(-\frac{\gamma}{\beta_r\overline{\gamma}_{ri}}\right)}{\left(1+\kappa_2\gamma\right)^{N_r-1}\left(1+\kappa_3\gamma\right)^{N_s}}\sum_{m=1}^{N_e}\frac{1}{\Gamma\left(m\right)}\left(\frac{\gamma}{\beta_r\overline{\gamma}_{ri}}\right)^{m-1}\notag\\
&\times\sum_{n=0}^{N_e-m}{N_r-1\choose n}\left(\kappa_2\gamma\right)^{n}\sum_{l=0}^{N_e-m-n}{N_s\choose l}\left(\kappa_3\gamma\right)^{l},
\end{align}
where $\overline{\gamma}_{ri} = P_rd_{ri}^{-\eta}\sigma_{i2}^{-2}$, $\kappa_2 = \frac{1-\beta_r}{\beta_r\left(N_r-1\right)}$, and $\kappa_3 = \frac{P_sd_{ri}^{\eta}}{\beta_rP_rN_sd_{si}^{\eta}}$.

\newcounter{TempEqCnt}
\setcounter{TempEqCnt}{\value{equation}}
\setcounter{equation}{31}
\begin{figure*}[ht]
\begin{align}\label{t1_result_3}
\mathcal{J}_2 =&
\left(1+\kappa_2\tau_e\right)^{-\left(N_r-1\right)}
\int_0^{\infty}\int_{0}^{\pi}d_{si}\frac{\exp\left(-\psi\left(\theta\right)\right)}{\left(1+\frac{P_s \psi\left(\theta\right)}{N_sd_{si}^{\eta}\sigma_{i2}^2}\right)^{N_s}}
\sum_{m=1}^{N_e}\frac{1}{\Gamma\left(m\right)}\left(\psi\left(\theta\right)\right)^{m-1}\notag\\
&\times
\sum_{n=0}^{N_e-m}{N_r-1\choose n}\left(\kappa_2\tau_e\right)^n
\sum_{l=0}^{N_e-m-n}{N_s\choose l}\left(\frac{P_s\psi\left(\theta\right)}{N_sd_{si}^{\eta}\sigma_{i2}^2}\right)^ldd_{si}d\theta,
\end{align}
\hrule
\begin{align}\label{t1_result_4}
\mathcal{J}_3
=&\left(1+\kappa_1\tau_e\right)^{-\left(N_s-1\right)}\left(1+\kappa_2\tau_e\right)^{-\left(N_r-1\right)}
\int_{0}^{\infty}\int_{0}^{\pi}d_{si}\exp\left(-\frac{\tau_e\sigma_{i1}^2}{\beta_s{}P_s}d_{si}^{\eta}\right)
\sum_{p=1}^{N_e}\frac{1}{\Gamma\left(p\right)}\left(\frac{\tau_e\sigma_{i1}^2}{\beta_s P_s}d_{si}^{\eta}\right)^{p-1}\sum_{q=0}^{N_e-p}{N_s-1\choose q}\left(\kappa_1\tau_e\right)^q\notag\\
&\times\frac{\exp\left(-\psi\left(\theta\right)\right)}{\left(1+\frac{P_s\psi\left(\theta\right)}{N_sd_{si}^{\eta}\sigma_{i2}^2}\right)^{N_s}}
\sum_{m=1}^{N_e}\frac{1}{\Gamma\left(m\right)}\left(\psi\left(\theta\right)\right)^{m-1}
\sum_{n=0}^{N_e-m}{N_r-1\choose n}\left(\kappa_2\tau_e\right)^n
\sum_{l=0}^{N_e-m-n}{N_s\choose l}\left(\frac{P_s\psi\left(\theta\right)}{N_sd_{si}^{\eta}\sigma_{i2}^2}\right)^ldd_{si}d\theta.
\end{align}
\hrule
\end{figure*}
\setcounter{equation}{\value{TempEqCnt}}

\subsection{Outage Probabilities}

In this subsection, we define the transmission outage event and the secrecy outage event and then characterize their probabilities. We first denote $C_b$ as the instantaneous capacity between the source and the destination. According to \eqref{snr_d}, $C_b$ is given by
\begin{align}\label{c_b}
C_b=\frac{1}{2}\log_2\left(1+\Gamma_D\right),
\end{align}
where the presence of the factor $1/2$ is due to the fact that two time slots are used in the transmission. We also denote $C_e$ as the instantaneous capacity between the source and the eavesdropper. According to \eqref{snr_e}, $C_e$ is given by
\begin{align}\label{c_e}
C_e=\frac{1}{2}\log_2\left(1+\Gamma_E\right).
\end{align}
We assume that the wiretap code is adopted in the transmission. We denote $\left(R_b,R_e\right)$ as the parameter pair for the adopted wiretap code, where $R_b$ denotes the transmission rate of the wiretap code, and $R_e$ denotes the redundancy rate of the wiretap code revealing the cost of preventing eavesdropping. We also assume that the source and the relay use the same $\left(R_b,R_e\right)$ to transmit, but with different codebooks. As such, we define that the transmission outage event occurs when $C_b<R_b$. In this event, the received signals at the destination are not reliably decoded. We also define that the secrecy outage event occurs when $C_e\geq R_e$. In this event, the eavesdropper is able to decode the transmitted signals and secrecy is compromised.

Based on the definition of the transmission outage event, we define the transmission outage probability as the probability that the equivalent instantaneous SNR at the destination is less than
$\tau_b = 2^{R_b}-1$. Mathematically, $P_{to}$ is formulated as
\begin{align}\label{p_to}
P_{to}={\Pr}\left(\Gamma_{D}<\tau_b\right).
\end{align}
Using \eqref{snr_d}, \eqref{cdf_1}, and \eqref{cdf_2}, we re-express the transmission outage probability in \eqref{p_to} as
\begin{align}\label{p_tr_2}
P_{to}&={\Pr}\left(\min\left\{\gamma_{sr},\gamma_{rd}\right\}<\tau_b\right)\notag\\
&=1-\left(1-F_{\gamma_{sr}}\left(\tau_b\right)\right)\left(1-F_{\gamma_{rd}}\left(\tau_b\right)\right)\notag\\
&=\frac{\det\left(\mathbf{\Xi}\left(\frac{\tau_b}{\beta_s\overline{\gamma}_{sr}}\right)\right)}{\Gamma_{u_1}\left(u_1\right)\Gamma_{v_1}\left(u_1\right)}
+\frac{\det\left(\mathbf{\Theta}\left(\frac{\tau_b}{\beta_r\overline{\gamma}_{rd}}\right)\right)}{\Gamma_{u_2}\left(u_2\right)\Gamma_{v_2}\left(u_2\right)}\notag\\
&\hspace{0.4cm}-\frac{\det\left(\mathbf{\Xi}\left(\frac{\tau_b}{\beta_s\overline{\gamma}_{sr}}\right)\right)\det\left(\mathbf{\Theta}\left(\frac{\tau_b}{\beta_r\overline{\gamma}_{rd}}\right)\right)}{\Gamma_{u_1}\left(u_1\right)\Gamma_{v_1}\left(u_1\right)\Gamma_{u_2}\left(u_2\right)\Gamma_{v_2}\left(u_2\right)}.
\end{align}

Based on the definition of the secrecy outage event, we define the secrecy outage probability as the probability that $\Gamma_{E}$ is larger than $\tau_e=2^{R_e}-1$. Mathematically, $P_{so}$ is formulated as
\begin{align}\label{p_sec}
P_{so}={\Pr}\left(\Gamma_{E}>\tau_e\right).
\end{align}
According to \eqref{snr_e}, \eqref{cdf_3}, and \eqref{cdf_4}, we derive an easy-to-compute expression for the secrecy outage probability in the following theorem.

\begin{theorem}\label{t1}
The secrecy outage probability of the relay wiretap channel is derived as
\begin{align}\label{t1_result}
P_{so}=1-\exp\left(-2\lambda\left(\mathcal{J}_1 + \mathcal{J}_2 - \mathcal{J}_3\right)\right),
\end{align}
where
\begin{align}\label{t1_result_2}
\mathcal{J}_{1}=&\frac{\pi}{\eta}\left(\frac{\beta_s
P_s}{\tau_e\sigma_{i1}^2}\right)^{\frac{2}{\eta}}
\left(1+\kappa_1\tau_e\right)^{-\left(N_s-1\right)}\notag\\
&\times\sum_{p=1}^{N_e}\frac{\Gamma\left(\frac{2}{\eta}+p-1\right)}{\Gamma\left(p\right)}\sum_{q=0}^{N_e-p}{N_s-1\choose q}\left(\kappa_1 \tau_e\right)^q,
\end{align}
$\mathcal{J}_2$ and $\mathcal{J}_3$ are given by \eqref{t1_result_3} and \eqref{t1_result_4}, respectively, shown at the top of the next page. In \eqref{t1_result_3} and \eqref{t1_result_4}, we have $\psi\left(\theta\right)=\frac{\tau_e\sigma_{i2}^2}{\beta_rP_r}\left(d_{sr}^2+d_{si}^2-2d_{sr}d_{si}\cos\theta\right)^{\frac{\eta}{2}}$.
\end{theorem}
\begin{IEEEproof}
See Appendix \ref{App_t1}.
\end{IEEEproof}

We find that \emph{Theorem \ref{t1}} provides an easy-to-compute tool for efficiently evaluating the secrecy outage probability. Although $\mathcal{J}_2$ and $\mathcal{J}_3$ for general $\eta$ cannot be obtained in closed-form, they can be easily calculated since only a double integral is involved in $\mathcal{J}_2$ and $\mathcal{J}_3$.

\subsection{Asymptotic Outage Probabilities}

In this subsection, we examine the asymptotic behavior of the outage probabilities as $N_s\to\infty$. The obtained asymptotic results are particular valuable for large-scale MIMO systems where the source (or equivalently, the BS) is equipped with a sufficiently large number of antennas. We first present the expression for the asymptotic transmission outage probability in the following corollary. \begin{corollary}\label{c1}
The asymptotic transmission outage probability when $N_s\to\infty$ is given by
\setcounter{equation}{33}
\begin{align}\label{c1_result}
P_{to}^{\infty}=\frac{\det\left(\mathbf{\Theta}\left(\frac{\tau_b}
{\beta_r\overline{\gamma}_{rd}}\right)\right)}{\Gamma_{u_2}\left(u_2\right)
\Gamma_{v_2}\left(u_2\right)}.
\end{align}
\end{corollary}
\begin{IEEEproof}
We express the asymptotic transmission outage probability when $N_s\to\infty$ as
\begin{align}\label{c1_proof}
P_{to}^{\infty}=\lim_{N_s\to\infty}P_{to}.
\end{align}
We note that
\begin{align}\label{c1_proof_2} \lim_{N_s\to\infty}\frac{\mathbf{\Xi}\left(\frac{\tau_b}{\beta_s\overline{\gamma}_{sr}}\right)}{\Gamma_{u_1}\left(u_1\right)\Gamma_{v_1}\left(u_1\right)}=0.
\end{align}
Substituting \eqref{c1_proof_2} into \eqref{c1_proof} yields the result.
\end{IEEEproof}

According to \emph{Corollary \ref{c1}}, we find that the asymptotic transmission outage probability is solely determined by $\overline{\gamma}_{rd}$ when $N_s\to\infty$. This finding is due to the fact that $\gamma_{sr}\to\infty$ when $N_s\to\infty$. As such, we conclude that the probability that $\Gamma_{D}$ is less than $\tau_b$ when $N_s\to\infty$ is determined by the link quality of the relay-destination channel only.

We next present the asymptotic secrecy outage probability when $N_s\to\infty$ in the following corollary.
\begin{corollary}\label{c2}
The asymptotic secrecy outage probability when $N_s\to\infty$ is given by
\begin{align}\label{c2_result}
P_{so}^{\infty}=1-\exp\left(-2\lambda\left(\mathcal{J}_1^{\infty}
+\mathcal{J}_2^{\infty}-\mathcal{J}_3^{\infty}\right)\right),
\end{align}
where
\begin{align}\label{c2_result_2}
\mathcal{J}_1^{\infty} = & \frac{\pi}{\eta}\left(\frac{\beta_{s}P_{s}}{\tau_e\sigma_{i1}^2}\right)^{\frac{2}{\eta}}
\exp\left(-\frac{1-\beta_s}{\beta_s}\tau_e\right)\notag\\
&\times\sum_{p=1}^{N_e}\frac{\Gamma\left(\frac{2}{\eta}+p-1\right)}{\Gamma\left(p\right)}
\sum_{q=0}^{N_e-p}\frac{\left(\frac{1-\beta_s}{\beta_s}\tau_e\right)^q}{\Gamma\left(q+1\right)},
\end{align}
$\mathcal{J}_2^{\infty}$ and $\mathcal{J}_3^{\infty}$ are given by \eqref{c2_result_3} and \eqref{c2_result_4}, respectively, shown at the top of the next page.
\end{corollary}
\begin{IEEEproof}
We express the asymptotic secrecy outage probability when $N_s\to\infty$ as
\setcounter{equation}{40}
\begin{align}\label{c2_proof_1}
P_{so}^{\infty}=\lim_{N_s\to\infty}P_{so}.
\end{align}
We note that
\begin{align}\label{c2_proof_2} \lim_{N_s\to\infty}\!\left(1+\frac{\left(1-\beta_s\right)\tau_e}{\beta_s\left(N_s-1\right)}\right)^{-\left(N_s-1\right)}=\exp\left(-\frac{1-\beta_s}{\beta_s}\tau_e\right),
\end{align}
and
\begin{align}\label{c2_proof_3}
\lim_{N_s\to\infty}{N_s-1\choose q}\left(\frac{\left(1-\beta_s\right)\tau_e}{\beta_s\left(N_s-1\right)}\right)^q=\frac{\left(\frac{1-\beta_s}{\beta_s}\tau_e\right)^q}{\Gamma\left(q+1\right)}.
\end{align}
We also note
\begin{align}\label{c2_proof_4}
\lim_{N_s\to\infty}\left(1+\frac{P_s\psi\left(\theta\right)}{N_sd_{si}^{\eta}\sigma_{i2}^2}\right)^{-N_s}=\exp\left(-\frac{P_s\psi\left(\theta\right)}{d_{si}^{\eta}\sigma_{i2}^2}\right),
\end{align}
and
\begin{align}\label{c2_proof_5}
\lim_{N_s\to\infty}{N_s\choose l}\left(\frac{P_s\psi\left(\theta\right)}{N_sd_{si}^{\eta}\sigma_{i2}^2}\right)^{l}=\frac{\left(\frac{P_s\psi\left(\theta\right)}{d_{si}^{\eta}\sigma_{i2}^2}\right)^l}{\Gamma\left(l+1\right)}.
\end{align}
Substituting \eqref{c2_proof_2}, \eqref{c2_proof_3}, \eqref{c2_proof_4}, and \eqref{c2_proof_5} into \eqref{c2_proof_1} yields \eqref{c2_result}, which completes the proof.
\end{IEEEproof}

According to \emph{Corollary \ref{c2}}, we find that the asymptotic secrecy outage probability approaches a certain value that is independent of $N_s$ when $N_s\to\infty$. This reveals that adding extra transmit antennas at the source does not always decrease the secrecy outage probability.

\setcounter{TempEqCnt}{\value{equation}}
\setcounter{equation}{38}
\begin{figure*}[ht]
\begin{align}\label{c2_result_3}
\mathcal{J}_2^{\infty} =&
\left(1+\kappa_2\tau_e\right)^{-\left(N_r-1\right)}
\int_0^{\infty}\int_{0}^{\pi}d_{si}{\exp\left(-\psi\left(\theta\right)-\frac{P_s \psi\left(\theta\right)}{d_{si}^{\eta}\sigma_{i2}^2}\right)}
\sum_{m=1}^{N_e}\frac{1}{\Gamma\left(m\right)}\left(\psi\left(\theta\right)\right)^{m-1}\notag\\
&\times\sum_{n=0}^{N_e-m}{N_r-1\choose n}\left(\kappa_2\tau_e\right)^n
\sum_{l=0}^{N_e-m-n}\frac{\left(\frac{P_s\psi\left(\theta\right)}{d_{si}^{\eta}\sigma_{i2}^2}\right)^l}{\Gamma\left(l+1\right)}dd_{si}d\theta,
\end{align}
\hrule
\begin{align}\label{c2_result_4}
\mathcal{J}_3^{\infty}
=&\frac{\exp\left(-\frac{1-\beta_s}{\beta_s}\tau_e\right)}{\left(1+\kappa_2\tau_e\right)^{N_r-1}}
\int_{0}^{\infty}\int_{0}^{\pi}d_{si}\exp\left(-\frac{\tau_e\sigma_{i1}^2}{\beta_s{}P_s}d_{si}^{\eta}\right)
\sum_{p=1}^{N_e}\frac{1}{\Gamma\left(p\right)}\left(\frac{\tau_e\sigma_{i1}^2}{\beta_s P_s}d_{si}^{\eta}\right)^{p-1}\sum_{q=0}^{N_e-p}\frac{\left(\frac{1-\beta_s}{\beta_s}\tau_e\right)^q}{\Gamma\left(q+1\right)}\notag\\
&\times{\exp\left(-\psi\left(\theta\right)-\frac{P_s \psi\left(\theta\right)}{d_{si}^{\eta}\sigma_{i2}^2}\right)}
\sum_{m=1}^{N_e}\frac{1}{\Gamma\left(m\right)}\left(\psi\left(\theta\right)\right)^{m-1}
\sum_{n=0}^{N_e-m}{N_r-1\choose n}\left(\kappa_2\tau_e\right)^n
\sum_{l=0}^{N_e-m-n}\frac{\left(\frac{P_s\psi\left(\theta\right)}{d_{si}^{\eta}\sigma_{i2}^2}\right)^l}{\Gamma\left(l+1\right)}dd_{si}d\theta.
\end{align}
\hrule
\end{figure*}
\setcounter{equation}{\value{TempEqCnt}}

\subsection{Secrecy Throughput}

So far, we have derived the exact and the asymptotic transmission outage probability and secrecy outage probability of the considered relay wiretap channel. Our derived expressions are valid for given $R_b$, $R_e$, $\beta_s$, and $\beta_r$. A question then naturally arises: ``\textit{How do we determine the optimal $\left(R_b^{\ast\circ},R_e^{\ast\circ},\beta_s^{\ast\circ},\beta_r^{\ast\circ}\right)$ that achieves the maximum secrecy performance of this relay wiretap channel, under a secrecy outage probability constraint?}'' Our answer to this question demonstrates the usefulness of our analytical expressions in a wider sense. It shows how the expressions can be embedded and utilized in a complex optimization problem, presenting a solution that can be determined in a much faster manner than would otherwise be possible. A concrete example of this usefulness in an operational sense, would be the dynamic and real-time determination of the optimal system parameter settings for a given secrecy performance metric.

To answer the question, we utilize our derived expressions directly to characterize the secrecy performance of the relay wiretap channel. As a specific example, we consider the metric termed the \emph{secrecy throughput} introduced by \cite{Zhou10}. This performance metric quantifies the average confidential information rate when the source transmits. The secrecy throughput for the relay wiretap channel is given by
\begin{align}\label{throughput}
T_s=\left(R_b-R_e\right)\left(1-P_{to}\right).
\end{align}
The maximization problem is accordingly formulated as
\begin{subequations}\label{prob_form}
\begin{align}
\max_{R_b,R_e,\beta_s,\beta_r}\hspace{0.5cm}&T_s,\\
s.t.\hspace{0.5cm}P_{so}&\leq \varphi,\\ 0\leq& R_e\leq R_b,\\ 0<&\beta_s\leq1, 0<\beta_r\leq1.
\end{align}
\end{subequations}
In the following, we describe in detail how the maximum secrecy throughput is obtained by judiciously selecting the transmission parameters\footnote{We note that the maximum secrecy throughput is formally a local maximum, since we numerically search the transmission parameters that jointly maximize the secrecy throughput only within anticipated ranges.}. To this end, we solve the maximization problem in \eqref{prob_form} in two steps. First, we fix power allocation factors $\beta_s$ and $\beta_r$, and choose the wiretap code rates pair, $\left(R_b^{\ast},R_e^{\ast}\right)$, that maximizes the secrecy throughput. Accordingly, the maximum secrecy throughput achieved by $\left(R_b^{\ast},R_e^{\ast}\right)$ for given $\beta_s$ and $\beta_r$ is defined as $T_s^{\ast}$. Second, we choose the wiretap code rates as well as the power allocation factor,
$\left(R_b^{\ast\circ},R_e^{\ast\circ},\beta_s^{\ast\circ},\beta_r^{\ast\circ}\right)$,
that jointly maximizes $T_s$. The details of these two steps are presented as follows:

\subsubsection{ $\left(R_b^{\ast}, R_e^{\ast}\right)$ for  given $\beta_s$ and $\beta_r$}

The wiretap code rates pair, $\left(R_b^{\ast},R_e^{\ast}\right)$, that maximizes $T_s$ for given $\beta_s$ and $\beta_r$ is determined as
\begin{subequations}\label{prob_form_2}
\begin{align}\label{prob_form_2_a}
\left(R_b^{\ast},R_e^{\ast}\right)&~~=~~\argmax~ T_s,\\
\label{prob_form_2_b}s.t.\hspace{0.4cm}&\hspace{0.4cm}P_{so}\leq\varphi,\\
\label{prob_form_2_c} &\hspace{0.4cm}0\leq{}R_e\leq{}R_b.
\end{align}
\end{subequations}

Taking the first-order derivative of $P_{so}$ with respect to $R_e$, we confirm that $\partial P_{so}/\partial R_e<0$, which indicates that $P_{so}$ monotonically decreases as $R_e$ increases. As such, the value of $R_e^{\ast}$ satisfying \eqref{prob_form_2_b} is the value of $R_e^{\ast}$ that satisfies the secrecy outage probability constraint, i.e., $P_{so}\left(R_e^{\ast}\right)=\varphi$. We then confirm that $\partial P_{to}/\partial R_b>0$, which shows that $P_{to}$ monotonically increases as $R_e$ increases. As such, we note that $T_s\to0$ as $R_b$ increases such that $P_{to}\to 1$. Defining $R_b^{\max}$ as the value of $R_b$ that satisfies $P_{to}\left(R_b\right)=1$, we rewrite \eqref{prob_form_2} as
\begin{subequations}
\label{prob_form_4}
\begin{align}
\left(R_b^{\ast},R_e^{\ast}\right)&~~=~~\argmax~ T_s,\\
s.t.\hspace{0.4cm}&\hspace{0.4cm}P_{so}\leq\varphi,\\
&\hspace{0.4cm}R_e^{\ast}\leq R_b\leq R_b^{\max}.
\end{align}
\end{subequations}
Although a closed-form solution for $\left(R_b^{\ast},R_e^{\ast}\right)$ is mathematically intractable, we are able to find the values of $\left(R_b^{\ast},R_e^{\ast}\right)$ in a numerical way.

\subsubsection{$\left(R_b^{\ast\circ},R_e^{\ast\circ},\beta_s^{\ast\circ},\beta_r^{\ast\circ}\right)$}

The wiretap code rates and power allocation factors which jointly maximizes $T_s$ in \eqref{throughput}, $\left(R_b^{\ast\circ},R_e^{\ast\circ},\beta_s^{\ast\circ},\beta_r^{\ast\circ}\right)$, is determined as
\begin{subequations}
\label{prob_form_3}
\begin{align}
\left(R_b^{\ast\circ}, R_e^{\ast\circ},\beta_s^{\ast\circ},\beta_r^{\ast\circ}\right)&~~=~~\argmax~ T_s,\\
s.t.\hspace{0.4cm}&\hspace{0.4cm}P_{so}\leq\varphi, \\
&\hspace{0.4cm}0\leq R_e\leq R_b, \\
&\hspace{0.4cm}0<\beta_s\leq1, 0<\beta_r\leq1.
\end{align}
\end{subequations}
Using \eqref{p_tr_2} and \eqref{t1_result}, we are able to solve \eqref{prob_form_3} numerically. Specifically, we first select the value of $\left(R_b^{\ast},R_e^{\ast}\right)$ satisfying \eqref{prob_form_4} for each value of $\beta_s$ and $\beta_r$. This leads to the secrecy throughput with $\left(R_b^{\ast},R_e^{\ast}\right)$, denoted by $T_s^{\ast}=\left(R_b^{\ast}-R_e^{\ast}\right)\left(1-P_{to}\right)$. We then select the value of $\beta_s^{\ast\circ}$ and the value of $\beta_r^{\ast\circ}$ that maximize $T_s^{\ast}$ for $0<\beta_s\leq 1$ and $0<\beta_r\leq 1$. Accordingly, the value of $\left(R_b^{\ast},R_e^{\ast}\right)$ associated with $\beta_s^{\ast\circ}$ and $\beta_r^{\ast\circ}$ is defined as $\left(R_b^{\ast\circ},R_e^{\ast\circ}\right)$. Finally, the maximum secrecy throughput achieved by $\left(R_b^{\ast\circ}, R_e^{\ast\circ},\beta_s^{\ast\circ}, \beta_r^{\ast\circ}\right)$ is defined as $T_s^{\ast\circ}$.

We note that our expressions for the outage probabilities can also be implemented in other performance metrics. For example, we can utilize our expressions to characterize the average secrecy rate of the relay wiretap channels in the presence of spatially random eavesdroppers.

\section{Numerical Results}\label{sec:numerical}

In this section, we present numerical results to validate our analysis of the outage probabilities. We also examine the impact of transmission parameters (e.g., $R_b$, $\beta_s$, $\beta_r$) and system parameters (e.g., $N_s$ and $\lambda$) on the secrecy throughput of the relay wiretap channel. Throughout this section we concentrate on the practical example of a highly shadowed urban area with $\eta = 4$.

\begin{figure}[t!]
\begin{center}{\includegraphics[height=3.2in,width = 3.3in]{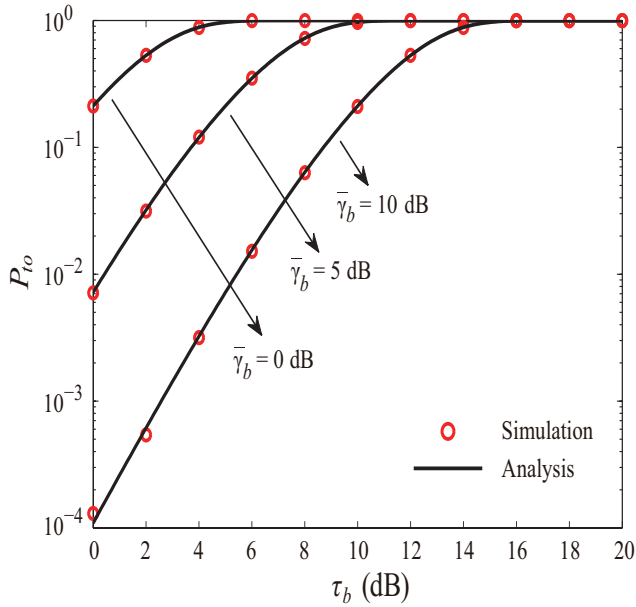}}
\caption{$P_{to}$ versus $\tau_b$ for different values of
$\overline{\gamma}_b$ with $\eta=4$, $N_s=4$, $N_r = 2$, $N_d = 2$, $N_e= 2$, $\beta_s = \beta_r =
0.5$, $\lambda = 0.01$, and $d_{sr} = d_{rd} = 10$.}\label{fig_side_a}
\end{center}
\end{figure}

We first demonstrate the accuracy of the transmission outage probability and the secrecy outage probability using Monte Carlo simulations. In Fig. \ref{fig_side_a}, we plot $P_{to}$ versus $\tau_b$ for different values of $\overline{\gamma}_b$ with $N_s=4$, $N_r = 2$, $N_d=2$, $N_e=2$, $\beta_s=\beta_r=0.5$, $\lambda = 0.01$, and $d_{sr} = d_{rd} = 10$. In this figure, we consider
$\overline{\gamma}_{sr}=\overline{\gamma}_{rd}=\overline{\gamma}_b$. We first see that the analytical curves, generated from \eqref{p_tr_2}, precisely match the simulation points marked by red circles, which demonstrates the correctness of our expression for $P_{to}$ in \eqref{p_tr_2}. Second, we see that $P_{to}$ increases monotonically as $\tau_b$ increases for a given $\overline{\gamma}_b$. This reveals that the transmission outage probability increases when the transmission rate of the wiretap code increases. We further see that $P_{to}$ decreases as $\overline{\gamma}_b$ increases for a given $\tau_b$. This reveals that the transmission outage probability reduces when the source and the relay use a higher power to transmit for a fixed $\tau_b$.

\begin{figure}[t!]
\begin{center}{\includegraphics[height=3.2in,width = 3.3in]{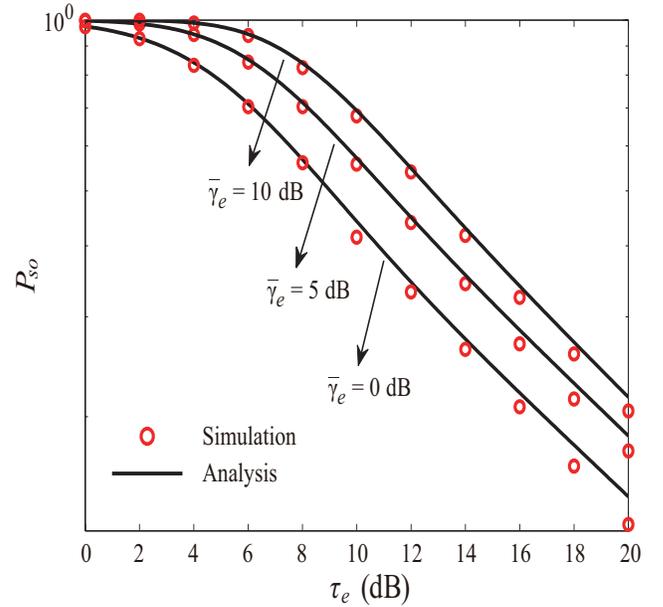}}
\caption{$P_{so}$ versus $\tau_e$ for different values of
$\overline{\gamma}_e$ with $\eta=4$, $N_s=4$, $N_r = 2$, $N_d = 2$, $N_e= 2$, $\beta_s = \beta_r=
0.5$, $\lambda = 0.01$, and $d_{sr} = d_{rd} = 10$.}\label{fig_side_b}
\end{center}
\end{figure}

In Fig. \ref{fig_side_b}, we plot $P_{so}$ versus $\tau_e$ for different values of $\overline{\gamma}_e$ with $N_s=4$, $N_r = 2$, $N_d=2$, $N_e=2$, and $\beta_s = \beta_r=0.5$. In
this figure we consider $\overline{\gamma}_{sr}{\sigma_r^2}/{\sigma_{i1}^2}=\overline{\gamma}_{rd}{\sigma_d^2}/{\sigma_{i2}^2}=\overline{\gamma}_e$. We see an excellent match between the analytical curves generated from \eqref{t1_result} and the simulation points marked by red circles, demonstrating the correctness of our expression for $P_{so}$ in \eqref{t1_result}. We then see that $P_{so}$ decreases monotonically as $\tau_e$ increases for a given $\overline{\gamma}_e$, which shows that the secrecy outage probability decreases when the redundancy rate of the wiretap code increases. We further observe that $P_{so}$ increases as $\overline{\gamma}_e$ increases. This is due to the fact the eavesdroppers receive signals from {\it both} the source and the relay. It follows that increasing the transmit power at the source and the relay leads to an improved received SNR at the eavesdroppers.

\begin{figure}[t!]
\begin{center}{\includegraphics[height=3.2in,width = 3.3in]{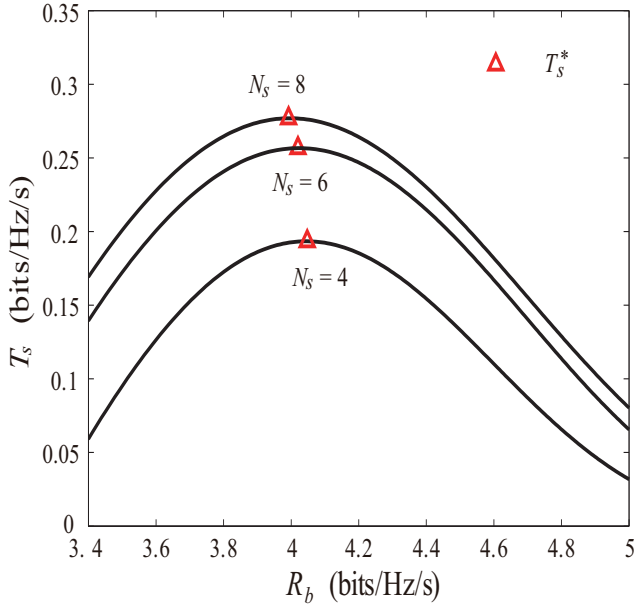}}
\caption{$T_s$ versus $R_b$ for different values of $N_s$ with $\eta=4$, $N_r = 2$, $N_d = 2$, $N_e= 2$, $\beta_s = \beta_r=0.5$, $\lambda = 0.01$, $d_{sr} = d_{rd} = 10$, $\varphi = 0.4$, $\overline{\gamma}_b = 10$ dB, and $\overline{\gamma}_b/\overline{\gamma}_e=20$.}\label{fig_side_c}
\end{center}
\end{figure}

In the following, we examine the impact of the transmission parameters on the secrecy throughput that is characterized by the derived outage probabilities. We first examine the impact of $R_b$ on $T_{s}$. In Fig. \ref{fig_side_c}, we plot $T_s$ versus $R_b$ for different values of $N_s$ with $R_e^{\ast}$ and fixed $N_d$, $N_e$, $\beta_s$, and $\beta_r$. We first observe that there exists a unique $R_b^{\ast}$ that maximizes $T_s$ for given $\beta_s$ and $\beta_r$. We also observe that the maximum $T_s$ for given $\beta_s$ and $\beta_r$, i.e., $T_s^{\ast}$, increases as $N_s$ increases. This shows that adding extra transmit antennas at the source significantly enhances the secrecy performance of the relay wiretap channel.

\begin{figure}[t!]
\begin{center}{\includegraphics[height=3.2in,width = 3.3in]{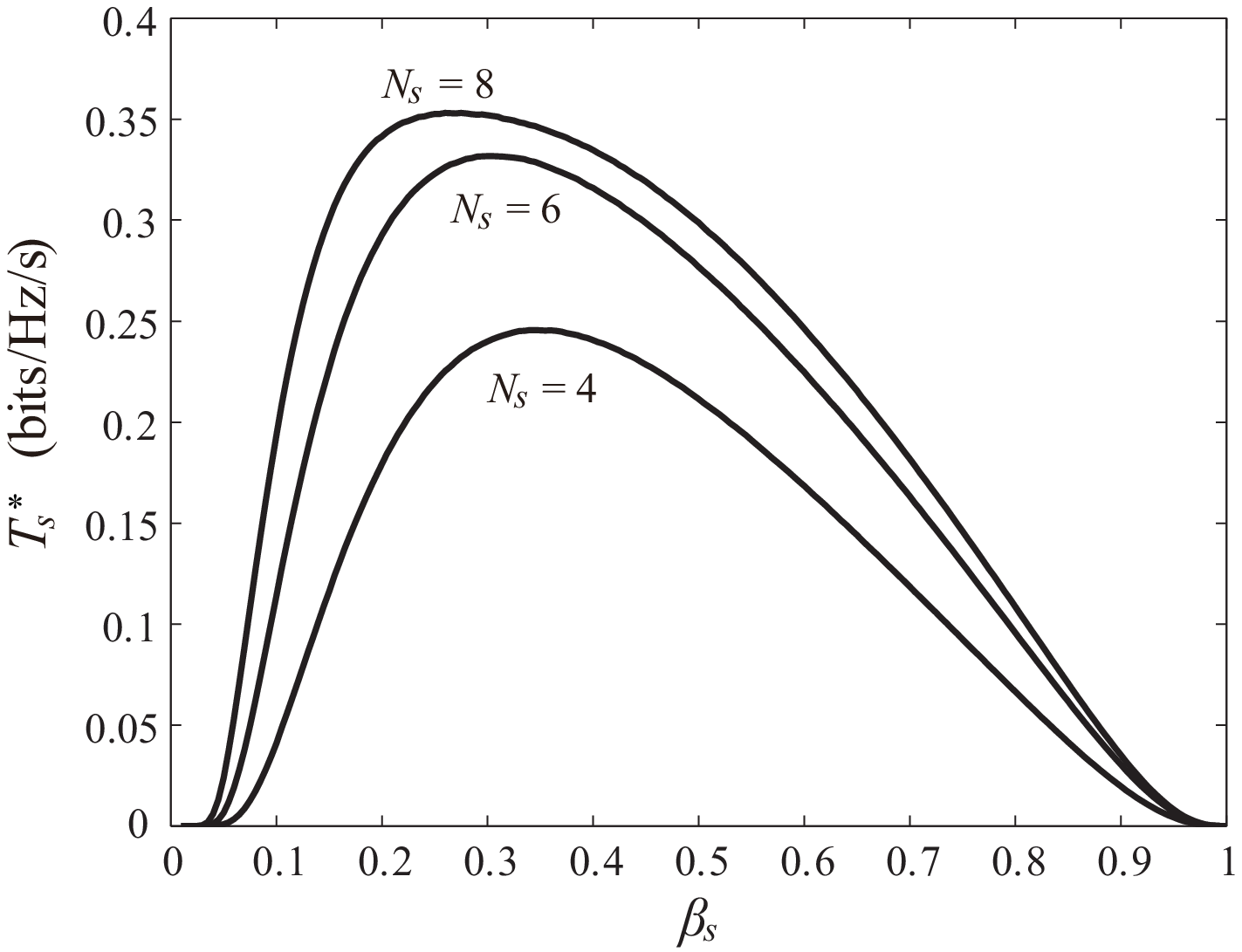}}
\caption{$T_s^{\ast}$ versus $\beta_s$ for different values of $N_s$
with $\eta=4$, $N_r = 2$, $N_d = 2$, $N_e= 2$, $\beta_r = 0.5$, $\lambda = 0.01$, $d_{sr} = d_{rd} = 10$, $\varphi = 0.4$, $\overline{\gamma}_b = 10$ dB, and $\overline{\gamma}_b/\overline{\gamma}_e=20$.}\label{fig_side_d}
\vspace{-0.1in}
\end{center}

\end{figure}

We next examine the impact of $\beta_s$ and $\beta_r$ on $T_{s}^{\ast}$. In Fig. \ref{fig_side_d}, we plot $T_s^{\ast}$ versus $\beta_s$ for different values of $N_s$ with a fixed $\beta_r$. For each point of $T_s^{\ast}$, we choose $\left(R_b^{\ast},R_e^{\ast}\right)$ that maximizes $T_s$ for the corresponding $\beta_s$. We first observe that there exists a unique $\beta_s^{\ast\circ}$ that maximizes $T_s^{\ast}$. We then observe that the maximum $T_s^{\ast}$ for a fixed $\beta_r$ increases as $N_s$ increases. Furthermore, we observe that the value of $\beta_s^{\ast\circ}$ slightly decreases as $N_s$ increases, which shows that in order to maintain the maximum secrecy throughput, the power allocated to AN signals at the source needs to be increased as the number of antennas at the source increases.

\begin{figure}[t!]
\begin{center}{\includegraphics[height=3.2in,width = 3.3in]{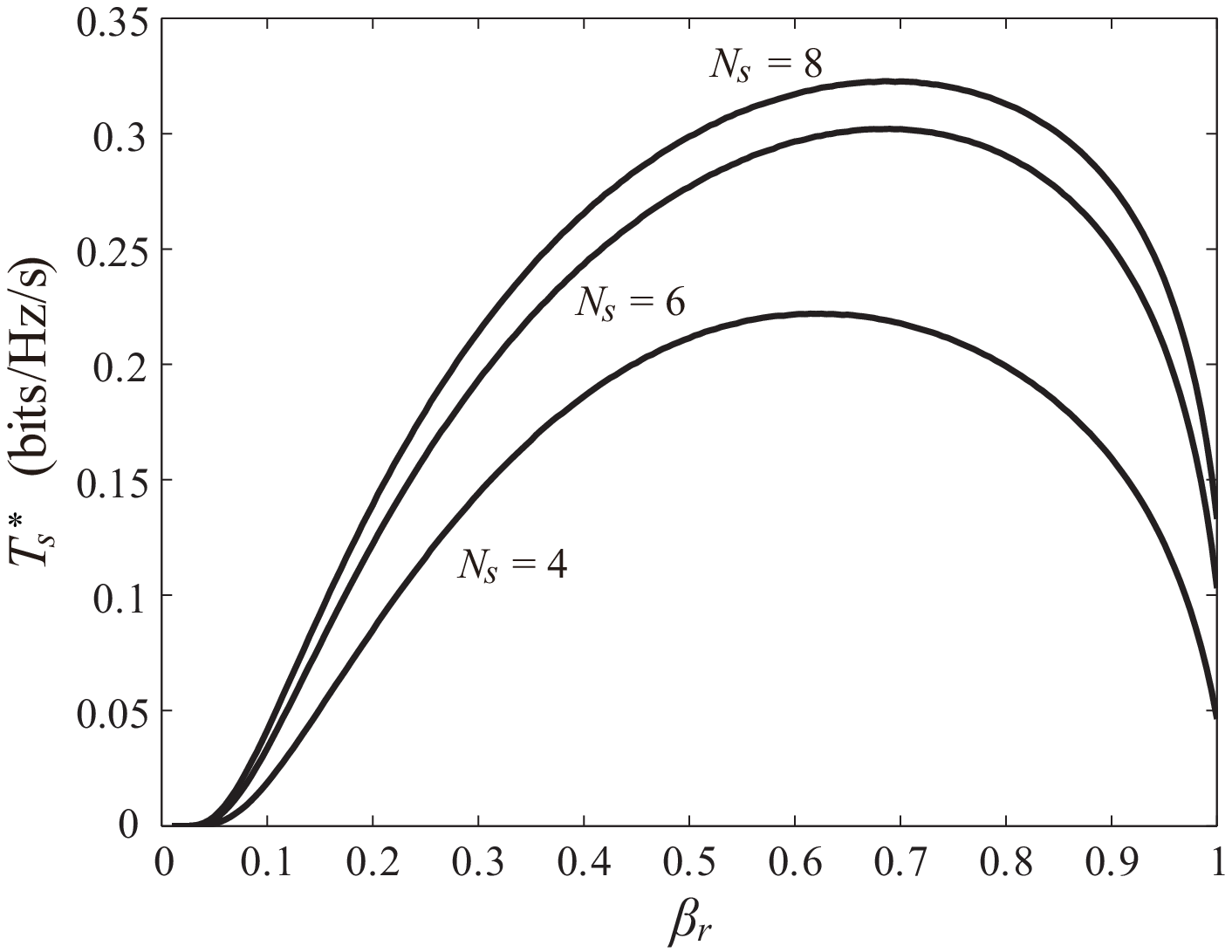}}
\caption{$T_s^{\ast}$ versus $\beta_r$ for different values of $N_s$
with $\eta=4$, $N_r = 2$, $N_d = 2$, $N_e= 2$, $\beta_s = 0.5$, $\lambda = 0.01$, $d_{sr} = d_{rd} = 10$, $\varphi = 0.4$, $\overline{\gamma}_b = 10$ dB, and
$\overline{\gamma}_b/\overline{\gamma}_e = 20$.}\label{fig_side_e}
\end{center}
\end{figure}

In Fig. \ref{fig_side_e}, we plot $T_s^{\ast}$ versus $\beta_r$ for different values of $N_s$ with a fixed $\beta_s$. Similar as Fig. \ref{fig_side_d}, for each point of $T_s^{\ast}$, we choose $\left(R_b^{\ast},R_e^{\ast}\right)$ that maximizes $T_s$ for the corresponding $\beta_r$. First, we see a unique $\beta_r^{\ast\circ}$ that maximizes $T_s^{\ast}$. Second, we see that the maximum $T_s^{\ast}$ for a fixed $\beta_s$ increases as $N_s$ increases. Additionally, we note that the value of $\beta_r^{\ast\circ}$ increases as $N_s$ increases, demonstrating that a lower power is needed to be allocated to AN signals at the relay in order to achieve the maximum secrecy throughput when the antenna number at the source increases.

\begin{figure}[t!]
\begin{center}{\includegraphics[height=3.2in,width = 3.3in]{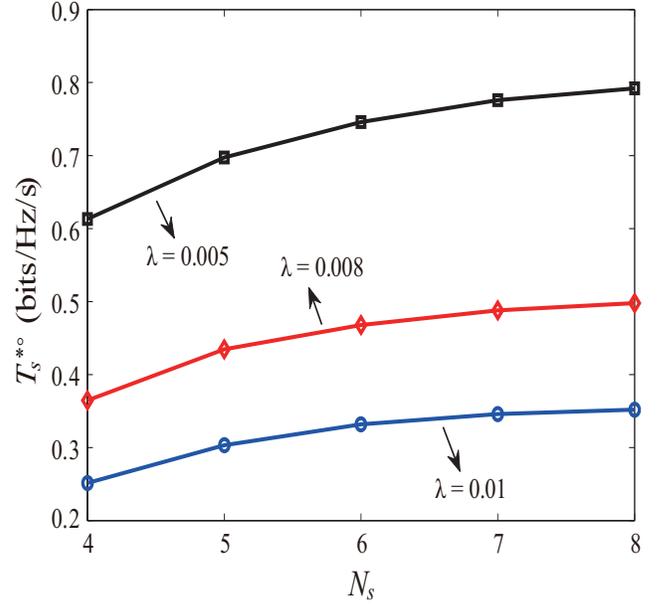}}
\caption{$T_s^{\ast\circ}$ versus $N_s$ for different values of $\lambda$
with $\eta=4$, $N_r = 2$, $N_d = 2$, $N_e= 2$, $d_{sr} = d_{rd} = 10$, $\varphi = 0.4$, $\overline{\gamma}_b = 10$ dB, and
$\overline{\gamma}_b/\overline{\gamma}_e = 20$.}\label{fig_side_f}
\end{center}
\end{figure}

Finally, we examine the impact of $N_{s}$ and $\lambda$ on $T_s^{\circ\ast}$. In Fig. \ref{fig_side_f}, we plot $T_s^{\ast\circ}$ versus $N_s$ for different values of $\lambda$. For each point of $T_s^{\ast\circ}$, we choose $\left(\beta_s^{\ast\circ},\beta_r,R_b^{\ast\circ},R_e^{\ast\circ}\right)$ that maximizes $T_s$. We first observe that $T_s^{\ast\circ}$ increases as $N_s$ increases. This observation is consistent with the observation in Fig. \ref{fig_side_c}. We then observe that $T_s^{\ast\circ}$ decreases as $\lambda$ increases. This is due to the fact that more eavesdroppers exist as $\lambda$ increases. The increasing number of eavesdroppers increases the value of $R_e^{\ast}$ that satisfies the secrecy constraint.


\section{Conclusion}\label{sec:conclusion}

We designed the secure transmission that maximizes the secrecy throughput of the generalized relay wiretap channel in the presence of spatially random multi-antenna eavesdroppers. In the transmission we assumed that both the source and the relay transmit AN signals with information signals in order to confuse the eavesdroppers. Considering the use of the decode-and-forward relaying protocol, we first derived a closed-form expression for the transmission outage probability and an easy-to-compute expression for the secrecy outage probability. We then derived simple yet valuable expressions for the asymptotic transmission outage probability and the asymptotic secrecy outage probability when the number of antennas at the source becomes sufficiently large. Using our derived expressions, we characterized the secrecy throughput of the the relay wiretap channel and then determined the transmission and system parameters that achieve the maximum secrecy throughput. Finally, we evaluated the impact of these parameters on the secrecy throughput.

\begin{appendices}

\section{Proof of Theorem \ref{t1}}\label{App_t1}

According to \eqref{snr_e}, \eqref{cdf_3}, and \eqref{cdf_4}, we re-express \eqref{p_sec} as
\setcounter{equation}{47}
\begin{align}\label{p_sec_2}
P_{so}&=1-{\Pr}\left\{\Gamma_{E}\leq\tau_e\right\}\notag\\
&=1-{\Pr}\left\{\max_{i\in\Phi} \left\{\max\left\{\gamma_{si},\gamma_{ri}\right\}\right\}\leq\tau_e\right\}\notag\\
&=1-\mathbb{E}_{\Phi}\left[\prod_{i\in\Phi}{\Pr}\left\{\max\left\{\gamma_{si},\gamma_{ri}\right\}\leq\tau_e\right\}\right]\notag\\
&=1-\mathbb{E}_{\Phi}\left[\prod_{i\in\Phi}F_{\gamma_{si}}\left(\tau_e\right)F_{\gamma_{ri}}\left(\tau_e\right)\right]\notag\\
&\overset{(a)}{=}1-\exp\left(-2\lambda\left(\mathcal{J}_1+\mathcal{J}_2-\mathcal{J}_3\right)\right).
\end{align}
where
\begin{align}\label{J_1}
\mathcal{J}_1 = &\int_0^{\infty}\!\!\int_{0}^{\pi}\!\!d_{si}\!\!\left(\!\frac{\exp\left(-\frac{\tau_e\sigma_{i1}^2}{\beta_s P_s}d_{si}^{\eta}\right)}{\left(1+\kappa_1\tau_e\right)^{N_s-1}}\!
\sum_{p=1}^{N_e}\frac{1}{\Gamma\left(p\right)}\left(\!\frac{\tau_e\sigma_{i1}^2}{\beta_s P_s}d_{si}^{\eta}\!\right)^{p-1}\right.\notag\\
&\hspace{1.5cm}\left.\times\sum_{q=0}^{N_e-p}{N_s-1\choose q}\left(\kappa_1\tau_e\right)^q
\right)dd_{si}d\theta,
\end{align}
\begin{align}\label{J_2}
&\mathcal{J}_2 \notag\\=&
\int_0^{\infty}\int_{0}^{\pi}d_{si}\frac{\exp\left(-\frac{\tau_e\sigma_{i2}^2}{\beta_rP_r}d_{ri}^{\eta}\right)}{\left(1+\kappa_2\tau_e\right)^{N_r-1}\left(1+\frac{P_s\tau_e d_{ri}^{\eta}}{\beta_rP_rN_sd_{si}^{\eta}}\right)^{N_s}}\notag\\
&\times\sum_{m=1}^{N_e}\frac{1}{\Gamma\left(m\right)}\left(\frac{\tau_e\sigma_{i2}^2}{\beta_rP_r}d_{ri}^{\eta}\right)^{m-1}\sum_{n=0}^{N_e-m}{N_r-1\choose n}\left(\kappa_2\tau_e\right)^n\notag\\
&\times\sum_{l=0}^{N_e-m-n}{N_s\choose l}\left(\frac{P_s\tau_e d_{ri}^{\eta}}{\beta_r P_rN_sd_{si}^{\eta}}\right)^ldd_{si}d\theta,
\end{align}
and
\begin{align}\label{J_3}
&\mathcal{J}_3\notag\\
=& \int_0^{\infty}\int_{0}^{\pi}d_{si}\frac{\exp\left(-\frac{\tau_e\sigma_{i1}^2}{\beta_s
P_s}d_{si}^{\eta}-\frac{\tau_e\sigma_{i2}^2}{\beta_rP_r}d_{ri}^{\eta}\right)}{\left(1+\kappa_1\tau_e\right)^{N_s-1}}\notag\\
&\times{\left(1+\kappa_2\tau_e\right)^{-\left(N_r-1\right)}\left(1+\frac{P_sd_{ri}^{\eta}\tau_e}{\beta_rP_rN_sd_{si}^{\eta}}\right)^{-N_s}}\notag\\
&\times\sum_{p=1}^{N_e}\frac{1}{\Gamma\left(p\right)}\left(\frac{\tau_e\sigma_{i1}}{\beta_s P_s}d_{si}^{\eta}\right)^{p-1}\sum_{q=0}^{N_e-p}{N_s-1\choose q}\left(\kappa_1\tau_e\right)^q\notag\\
&\times\sum_{m=1}^{N_e}\frac{1}{\Gamma\left(m\right)}\left(\frac{\tau_e\sigma_{i2}^2}{\beta_rP_r}d_{ri}^{\eta}\right)^{m-1}\sum_{n=0}^{N_e-m}{N_r-1\choose n}\left(\kappa_2\tau_e\right)^n\notag\\
&\times\sum_{l=0}^{N_e-m-n}{N_s\choose l}\left(\frac{P_sd_{ri}^{\eta}\tau_e}{\beta_r P_rN_sd_{si}^{\eta}}\right)^ldd_{si}d\theta.
\end{align}
In \eqref{p_sec_2}, the operation $(a)$ can be justified by applying the probability generating functional (PGFL) for the PPP $\Phi$, given by \cite{stoyan}
\begin{align}\label{pgfl}
\mathbb{E}_{\Phi}\left[\prod_{x\in\Phi}f\left(x\right)\right]=\exp\left\{-\int_{\mathbb{R}^2}\left[1-f\left(x\right)\right]\lambda
dx\right\},
\end{align}
and by changing to polar coordinates.

In order to proceed with our analysis we first derive $\mathcal{J}_1$ as
\begin{align}\label{J_1_2}
\mathcal{J}_1=&\pi\left(1+\kappa_1\tau_e\right)^{-\left(N_s-1\right)}\sum_{p=1}^{N_e}\frac{1}{\Gamma\left(p\right)}\sum_{q=0}^{N_e-p}{N_s-1\choose q}\left(\kappa_1\tau_e\right)^q\notag\\
&\times \int_{0}^{\infty}d_{si}\exp\left(-\frac{\tau_e\sigma_{i1}^2}{\beta_s P_s}d_{si}^{\eta}\right)\left(\frac{\tau_e\sigma_{i1}^2}{\beta_s P_s}d_{si}^{\eta}\right)^{p-1} dd_{si}\notag\\
\overset{(b)}{=}&\frac{\pi}{2}\left(1+\kappa_1\tau_e\right)^{-\left(N_s-1\right)}\sum_{p=1}^{N_e}\frac{1}{\Gamma\left(p\right)}\sum_{q=0}^{N_e-p}{N_s-1\choose q}\left(\kappa_1\tau_e\right)^q\notag\\
&\times\int_{0}^{\infty}\exp\left(-\frac{\tau_e\sigma_{i1}^2}{\beta_s P_s}u^{\frac{\eta}{2}}\right)\left(\frac{\tau_e\sigma_{i1}^2}{\beta_s P_s}u^{\frac{\eta}{2}}\right)^{p-1}du\notag\\
\overset{(c)}{=}&\frac{\pi}{\eta}\left(\frac{\beta_s P_s}{\tau_e\sigma_{i1}^2}\right)^{\frac{2}{\eta}}\left(1+\kappa_1\tau_e\right)^{-\left(N_s-1\right)}\notag\\
&\hspace{1cm}\times\sum_{p=1}^{N_e}\frac{1}{\Gamma\left(p\right)}\sum_{q=0}^{N_e-p}{N_s-1\choose q}\left(\kappa_1\tau_e\right)^q\notag\\
&\hspace{2.5cm}\times\int_0^{\infty}\exp\left(-t\right)t^{\frac{2}{\eta}+p-1}dt\notag\\
\overset{(d)}{=}&\frac{\pi}{\eta}\left(\frac{\beta_s
P_s}{\tau_e\sigma_{i1}^2}\right)^{\frac{2}{\eta}}
\left(1+\kappa_1\tau_e\right)^{-\left(N_s-1\right)}\notag\\
&\times\sum_{p=1}^{N_e}\frac{\Gamma\left(\frac{2}{\eta}+p-1\right)}{\Gamma\left(p\right)}\sum_{q=0}^{N_e-p}{N_s-1\choose q}\left(\kappa_1\tau_e\right)^q,
\end{align}
where in $(b)$ we use $u=d_{si}^2$, in $(c)$ we use $t=\frac{\tau_e\sigma_{i1}^2}{\beta_s P_s}u^{\frac{\eta}{2}}$, and $(d)$ follows from the definition of the gamma function. Similarly, we derive $\mathcal{J}_2$ as
\begin{align}\label{J_2_2}
&\mathcal{J}_2 \notag\\\overset{\left(e\right)}{=}&
\left(1+\kappa_2\tau_e\right)^{-\left(N_r-1\right)}\notag\\
&\times\int_0^{\infty}\int_{0}^{\pi}d_{si}\frac{\exp\left(-\frac{\tau_e\sigma_{i2}^2}{\beta_rP_r}\left(d_{sr}^2+d_{si}^2-2d_{sr}d_{si}\cos\theta\right)^{\frac{\eta}{2}}\right)}{\left(1+\frac{P_s\tau_e \left(d_{sr}^2+d_{si}^2-2d_{sr}d_{si}\cos\theta\right)^{\frac{\eta}{2}}}{\beta_rP_rN_sd_{si}^{\eta}}\right)^{N_s}}\notag\\
&\times\sum_{m=1}^{N_e}\frac{1}{\Gamma\left(m\right)}\left(\frac{\tau_e\sigma_{i2}^2}{\beta_rP_r}\left(d_{sr}^2+d_{si}^2-2d_{sr}d_{si}\cos\theta\right)^{\frac{\eta}{2}}\right)^{m-1}\notag\\
&\times\sum_{n=0}^{N_e-m}{N_r-1\choose n}\left(\kappa_2\tau_e\right)^n
\sum_{l=0}^{N_e-m-n}{N_s\choose l}\left(\frac{P_s\tau_e }{\beta_r P_rN_sd_{si}^{\eta}}\right)^l\notag\\
&\times\left(\left(d_{sr}^2+d_{si}^2-2d_{sr}d_{si}\cos\theta\right)^{\frac{\eta}{2}}\right)^ldd_{si}d\theta,
\end{align}
where in $\left(e\right)$ we apply the cosine formula. We further derive $\mathcal{J}_3$ as
\begin{align}\label{J_3_2}
&\mathcal{J}_3\notag\\
=&\left(1+\kappa_1\tau_e\right)^{-\left(N_s-1\right)}\left(1+\kappa_2\tau_e\right)^{-\left(N_r-1\right)}\notag\\
&\times\int_{0}^{\infty}\int_{0}^{\pi}d_{si}\exp\left(-\frac{\tau_e\sigma_{i1}^2}{\beta_s{}P_s}d_{si}^{\eta}\right)\notag\\
&\times\sum_{p=1}^{N_e}\frac{1}{\Gamma\left(p\right)}\left(\frac{\tau_e\sigma_{i1}^2}{\beta_s P_s}d_{si}^{\eta}\right)^{p-1}\sum_{q=0}^{N_e-p}{N_s-1\choose q}\left(\kappa_1\tau_e\right)^q\notag\\
&\times\frac{\exp\left(-\frac{\tau_e\sigma_{i2}^2}{\beta_rP_r}\left(d_{sr}^2+d_{si}^2-2d_{sr}d_{si}\cos\theta\right)^{\frac{\eta}{2}}\right)}{\left(1+\frac{P_s\tau_e \left(d_{sr}^2+d_{si}^2-2d_{sr}d_{si}\cos\theta\right)^{\frac{\eta}{2}}}{\beta_rP_rN_sd_{si}^{\eta}}\right)^{N_s}}\notag\\
&\times\sum_{m=1}^{N_e}\frac{1}{\Gamma\left(m\right)}\left(\frac{\tau_e\sigma_{i2}^2}{\beta_rP_r}\left(d_{sr}^2+d_{si}^2-2d_{sr}d_{si}\cos\theta\right)^{\frac{\eta}{2}}\right)^{m-1}\notag\\
&\times\sum_{n=0}^{N_e-m}{N_r-1\choose n}\left(\kappa_2\tau_e\right)^n
\sum_{l=0}^{N_e-m-n}{N_s\choose l}\left(\frac{P_s\tau_e }{\beta_r P_rN_sd_{si}^{\eta}}\right)^l\notag\\
&\times\left(\left(d_{sr}^2+d_{si}^2-2d_{sr}d_{si}\cos\theta\right)^{\frac{\eta}{2}}\right)^ldd_{si}d\theta,
\end{align}

Substituting \eqref{J_1_2}, \eqref{J_2_2}, and \eqref{J_3_2} into \eqref{p_sec_2}, we obtain the desired result in \eqref{t1_result}, which completes the proof.
\end{appendices}


\begin{thebibliography}{10}

\bibitem{Hong}
Y.-W. P. Hong, P.-C. Lan, and C.-C. J. Kuo,
\newblock ``Enhancing physical-layer secrecy in multiantenna wireless systems:
An overview of signal processing approaches,''
\newblock {\em IEEE Signal Process. Mag.}, vol. 30, no. 5, pp. 29--40, Sep. 2013.

\bibitem{Yang_Mag}
N. Yang, L. Wang, G. Geraci, M. Elkashlan, J. Yuan, and M. Di
Renzo,``Safeguarding 5G wireless communication networks using
physical layer security," {\em IEEE Commun. Mag.}, vol. 53, no. 4, pp. 20--27, Apr. 2015.

\bibitem{wyner}
A. Wyner,
\newblock ``The wire-tap channel,''
\newblock {\em Bell Syst. Tech. J.}, vol. 54, no. 8, pp. 1355--1387, Oct. 1975.

\bibitem{csiszar} I. Csisz\'{a}r and J. K\"{o}rner, ``Broadcast channels with confidential messages,'' {\em IEEE Trans. Inf. Theory}, vol. 24, no. 3, pp. 339--348, May 1978.

\bibitem{khisti}
A. Khisti and G. W. Wornell, ``Secure transmission with multiple
antennas \uppercase\expandafter{\romannumeral1}: The MISOME wiretap
channel,'' {\em IEEE Trans. Inf. Theory,} vol. 56, no. 6, pp.
3088--3104, Jul. 2010.

\bibitem{wornell}
A. Khisti and G. W. Wornell, ``Secure transmission with multiple
antennas \uppercase\expandafter{\romannumeral2}: The MIMOME wiretap
channel,'' {\em IEEE Trans. Inf. Theory,} vol. 56, no. 11, pp.
5515--5532, Nov. 2010.

\bibitem{chenxi}
C. Liu, G. Geraci, N. Yang, J. Yuan, and R. Malaney, ''Beamforming
for MIMO Gaussian channels with imperfect channel state
information,'' in {\em Proc. IEEE GlobeCOM 2013}, Atlanta, USA, Dec.
2013.

\bibitem{chenxi2}
C. Liu, N. Yang, G. Geraci, J. Yuan, and R. Malaney, ``Secrecy in
MIMOME wiretap channels: Beamforming with imperfect CSI,'' in {\em
Proc. IEEE ICC 2014}, Sydney, Australia, Jun. 2014.


\bibitem{nan4}
N. Yang, G. Geraci, J. Yuan, and R. Malaney, ``Confidential
broadcasting via linear precoding in non-homogeneous MIMO multiuser
networks,'' {\em IEEE Trans. Commun.}, vol. 62, no. 7, pp.
2515--2530, Jul. 2014.

\bibitem{Zhou10}
X. Zhou and M. R. McKay,
\newblock ``Secure transmission with artificial noise over fading channels: achievable rate and optimal power allocation,''
\newblock {\em IEEE Trans. Veh. Technol.}, vol. 59, no. 8, pp. 3831--3842, Oct. 2010.

\bibitem{Zhang13}
X. Zhang, X. Zhou, and M. R. McKay,
\newblock ``On the design of artificial-noise-aided secure multi-antenna transmission in slow fading channels,''
\newblock {\em IEEE Trans. Veh. Technol.}, vol. 62, no. 5, pp. 2170--2181, Jun. 2013.

\bibitem{nan5} N. Yang, S. Yan, J. Yuan, R. Malaney, R. Subramanian, and I. Land, ``Artificial noise: Transmission optimization in multi-input single-output wiretap channels,'' {\em IEEE Trans. Commun.}, vol. 63, no. 5, pp. 1771--1783, May 2015.

\bibitem{nan6} N. Yang, M. Elkashalan, T. Q. Duong, J. Yuan, and R. Malaney, ``Optimal transmission with artificial noise in MISOME wirtap channels,'' {\em IEEE Trans. Veh. Technol.}, accepted to appear.

\bibitem{nan}
N. Yang, P. L. Yeoh, M. Elkashlan, R. Schober, and I. B. Collings,
``Transmit antenna selection for security enhancement in MIMO
wiretap channels,'' {\em IEEE Trans. Commun.}, vol. 61, no. 1, pp.
144--154, Jan. 2013.

\bibitem{nan2}
N. Yang, H. A. Suraweera, I. B. Collings, and C. Yuen, ``Physical
layer security of TAS/MRC with antenna correlation,'' {\em IEEE
Trans. Inf. Foren. Sec.}, vol. 8, no. 1, pp. 254--259, Jan.
2013.

\bibitem{nan3}
N. Yang, P. L. Yeoh, M. Elkashlan, R. Schober, and J. Yuan, ``MIMO
wiretap channels: A secure transmission using transmit antenna
selection and receive generalized selection combining,'' {\em IEEE
Commun. Lett.}, vol. 17, no. 9, pp. 1754--1757, Sep. 2013.

\bibitem{shihao}
S. Yan, N. Yang, R. Malaney, and J. Yuan, ``Transmit antenna
selection with Alamouti coding and power allocation in MIMO wiretap
channels,'' {\em IEEE Trans. Wireless Commun.}, vol. 13, no. 3, pp.
1656--1667, Mar. 2014.


%


\bibitem{xiangyun_2}
X. Zhou, R. K. Ganti, J. G. Andrews, and A. Hj{\o}rungnes, ``On the
throughput cost of physical layer security in decentralized wireless
networks,'' {\em IEEE Trans. Wireless Commun.}, vol. 10, no. 8, pp.
2764--2775, Aug. 2011.

\bibitem{wang_he}
H. Wang, X. Zhou, and M. C. Reed, ``Physical layer security in
cellular networks: A stochastic geometry approach,'' {\em IEEE
Trans. Wireless Commun.}, vol. 12, no. 6, pp. 2776--2787, Jun. 2013.

\bibitem{gio_1}
G. Geraci, S. Singh, J. G. Andrews, J. Yuan, and I. B. Collings,
``Secrecy rates in broadcast channels with confidential messages and
external eavesdroppers,'' {\em IEEE Trans. Wireless Commun.}, vol.
13, no. 5, pp. 2931--2943, May 2014.

\bibitem{gio_2}
G. Geraci, H. S. Dhillon, J. G. Andrews, J. Yuan, and I. B.
Collings, ``Physical layer security in downlink multi-antenna
cellular networks,'' {\em IEEE Trans. Commun.}, vol. 62, no. 6, pp.
2006--2021, Jun. 2014.

\bibitem{huiming} T.-X. Zheng, H.-M. Wang, J. Yuan, D. Towsley, and M. H. Lee, ``Multi-antenna transmission with artificial noise against randomly distributed eavesdroppers,'' {\em IEEE Trans. Commun.}, accepted to appear.

\bibitem{Laneman}
J. N. Laneman, D. N. C. Tse, and G. W. Wornell, ``Cooperative
diversity in wireless networks: Efficient protocols and outage
behavior,'' {\em IEEE Trans. Inf. Theory}, vol. 50, pp. 3062--3080,
Dec. 2004.

\bibitem{Bassily}
R. Bassily, E. Ekrem, X. He, E. Tekin, J. Xie, M. R. Bloch, S.
Ulukus, and A. Yener,
\newblock ``Cooperative security at the physical layer: A summary of recent advances,''
\newblock {\em IEEE Signal Process. Mag.}, vol. 30, no. 5, pp. 16--28, Sep. 2013.

\bibitem{dong}
L. Dong, Z. Han, A. Petropulu, and H. V. Poor, ``Improving wireless
physical layer security via cooperating relays,'' {\em IEEE Trans.
Signal Process.}, vol. 58, no. 3, pp. 1875--1888, Mar. 2010.

\bibitem{huang} J. Huang, and A. Lee Swindlehurst, ``Cooperative jamming for secure communications in MIMO relay networks,'' {\em IEEE Trans. Signal Process.},
vol. 59, no. 10, pp. 4871--4884, Oct. 2011.

\bibitem{xiaoming_1} X. Chen, C. Zhong, C. Yuen, and H.-H. Chen, ``Multi-antenna relay aided wireless physical layer security,'' {\em IEEE Commun. Mag.}, accepted to appear.

\bibitem{xiaoming_2} X. Chen, L. Lei, H. Zhang, and C. Yuen, ``Large-scale MIMO relaying techniques for physical layer security: AF or DF?'' {\em IEEE Trans. Wireless Commun.}, accepted to appear.


\bibitem{hanzhu}
J. Chen, L. Song, Z. Han, and B. Jiao, ``Joint relay and jammer
selection for secure two-way relay networks,'' {\em IEEE Trans. Inf.
Foren. Sec.}, vol. 7, no. 1, pp. 310--320, Feb. 2012.

\bibitem{yulong}
Y. Zou, X. Wang, and W. Shen, ``Optimal relay selection for
physical-layer security in cooperative wireless networks,'' {\em
IEEE J. Sel. Areas Commun.}, vol. 31, no. 10, pp. 2099--2111, Oct.
2013.


\bibitem{lai} L. Lai and H. El Gamal, ``The relay-eavesdropper channel: Cooperation for secrecy,'' {\em IEEE Trans. Inf. Theory}, vol. 54, no. 9, pp. 4005--4019, Sep. 2008.

\bibitem{chenxi4}
C. Liu, N. Yang, J. Yuan, and R. Malaney, ``Location-based secure
transmission for wiretap channels,'' {\em IEEE J. Sel. Areas
Commun.}, vol. 33, no. 7, pp. 1458--1470, Jul. 2015.

\bibitem{chenxi5} C. Liu, N. Yang, J. Yuan, and R. Malaney, ``Secure transmission for relay wiretap channels in the presence of spatially random eavesdroppers,'' in {\em Proc. IEEE Globecom workshop on TCPLS}, San Diego, USA, Dec. 2015.




\bibitem{weber} S. Weber, J. G. Andrews, and N. Jindal, ``An overview of the transmission capacity of wireless networks,'' {\em IEEE Trans. Commun.}, vol. 58, no. 12, pp. 3593--3604, Dec. 2010.
\bibitem{dighe} P. A. Dighe, R. K. Mallik, and S. S. Jamuar, ``Analysis of transmit-recieve diversity in Rayleigh fading,'' {\em IEEE Trans. Commun.}, vol. 51, no.4, pp. 674--703, Apr. 2003.

\bibitem{MRC} M. Kang and M.-S. Alouini, ``A comparative study on the performance of MIMO MRC with and without cochannel interference,'' {\em IEEE Trans. Commun.}, vol. 52, no. 8, pp. 1417--1425, Aug. 2004.

\bibitem{mathew} M. R. Mckay, A. J. Grant, and I. B. Collings, ``Performance analysis of MIMO-MRC in double-correlated Rayleigh Environments,'' {\em IEEE Trans. Commun.}, vol. 55, no.3, pp. 497--507, Mar. 2007.


\bibitem{Dighe} P. A. Dighe, R. K. Mallik, and S. S. Jamuar, ``Analysis of transmit-receive diversity in Rayleigh fading,'' {\em IEEE Trans. Commun.}, vol. 51, no. 4, pp. 694--703, Apr. 2003.

\bibitem{table}
I. S. Gradshteyn and I. M. Ryzhik, {\em Tabel of Integrals, Series,
and Products,} 7th edition. Academic Press, 2007.

\bibitem{smith} H. Gao, P. J. Smith, and M. V. Clark, ``Theorectical reliability of MMSE linear diversity combining in Rayleigh-fading additive interference channels,'' {\em IEEE Trans. Commun.}, vol. 46, no. 5, pp. 666--672, May 1998.
%


\bibitem{stoyan}
D. Stoyan, W. Kendall, and J. Mecke, {\em Stochastic Geometry and
its Applications}, 2nd ed. John Wiley $\&$ Sons Ltd., 1996.
%

%
\end{thebibliography}
\end{document}